\documentclass[twoside,journey]{IEEEtran}
\usepackage{makecell}
\usepackage{array}
\usepackage{graphicx,amssymb,amsmath}
\usepackage{multicol}
\usepackage[noadjust]{cite}
\usepackage{setspace}
\usepackage{subfigure}
\usepackage{graphicx}
\usepackage{float}
\usepackage {url}
\usepackage{stfloats}
\usepackage{amsthm,pifont}
\usepackage{flushend}
\usepackage{cases,subeqnarray}
\usepackage{bm,multirow,bigstrut}
\usepackage{amsmath, amsthm, amssymb}
\usepackage{textcomp}
\usepackage{latexsym,bm}
\usepackage{booktabs}
\usepackage{xcolor}
\usepackage{mathtools}
\usepackage{dsfont}
\usepackage{extarrows}
\usepackage{epsfig}
\usepackage{epsfig}
\usepackage{epstopdf}
\usepackage[noend]{algpseudocode}
\usepackage{algorithmicx,algorithm}

\usepackage{cite}
\usepackage{bm}
\usepackage{cleveref}
\usepackage{multicol}       
\usepackage{multirow}       
\usepackage{array}          
\usepackage{colortbl}
\usepackage{makecell}
\definecolor{crimson}{RGB}{192,0,0}         
\definecolor{navy}{RGB}{47,85,151}         
\usepackage{bbding}
\usepackage{graphicx}
\usepackage{booktabs}
\usepackage{algorithm}
\usepackage{algpseudocode}

\theoremstyle{plain}

\theoremstyle{plain}

\usepackage{amsmath}

\IEEEoverridecommandlockouts
\begin{document}

\title{RIS-Aided Cell-Free Massive MIMO Systems for 6G: Fundamentals, System Design, and Applications}

\author{Enyu Shi, Jiayi Zhang,~\IEEEmembership{Senior Member,~IEEE}, Hongyang Du,
 Bo Ai,~\IEEEmembership{Fellow,~IEEE}, Chau Yuen,~\IEEEmembership{Fellow,~IEEE}, Dusit Niyato,~\IEEEmembership{Fellow,~IEEE}, Khaled B. Letaief,~\IEEEmembership{Fellow,~IEEE}, and Xuemin (Sherman) Shen,~\IEEEmembership{Fellow,~IEEE}

\thanks{E. Shi, J. Zhang, and B. Ai  are with the School of Electronic and Information Engineering, Beijing Jiaotong University, Beijing 100044, P. R. China. (e-mail: jiayizhang@bjtu.edu.cn).}
\thanks{H. Du and D. Niyato are with the School of Computer Science and Engineering, Nanyang Technological University, Singapore (e-mail: dniyato@ntu.edu.sg).}
\thanks{C. Yuen is with the School of Electrical and Electronics Engineering, Nanyang Technological University, Singapore 639798, Singapore (e-mail: chau.yuen@ntu.edu.sg).}
\thanks{Khaled B. Letaief is with the Department of Electrical and Computer Engineering, The Hong Kong University of Science and Technology, Hong Kong, and also with the Pengcheng Laboratory, Shenzhen 518055, Guangdong, China. (e-mail: eekhaled@ece.ust.hk).}
\thanks{X. Shen is with the Department of Electrical and Computer Engineering, University of Waterloo, Waterloo, ON N2L 3G1, Canada. (email:
sshen@uwaterloo.ca).}
}

\maketitle
\vspace{-1cm}
\begin{abstract}
An introduction of intelligent interconnectivity for people and things has posed higher demands and more challenges for sixth-generation (6G) networks, such as high spectral efficiency and energy efficiency, ultra-low latency, and ultra-high reliability. Cell-free (CF) massive multiple-input multiple-output (mMIMO) and reconfigurable intelligent surface (RIS), also called intelligent reflecting surface (IRS), are two promising technologies for coping with these unprecedented demands. Given their distinct capabilities, integrating the two technologies to further enhance wireless network performances has received great research and development attention. In this paper, we provide a comprehensive survey of research on RIS-aided CF mMIMO wireless communication systems. We first introduce system models focusing on system architecture and application scenarios, channel models, and communication protocols. Subsequently, we summarize the relevant studies on system operation and resource allocation, providing in-depth analyses and discussions. Following this, we present practical challenges faced by RIS-aided CF mMIMO systems, particularly those introduced by RIS, such as hardware impairments and electromagnetic interference. We summarize corresponding analyses and solutions to further facilitate the implementation of RIS-aided CF mMIMO systems. Furthermore, we explore an interplay between RIS-aided CF mMIMO and other emerging 6G technologies, such as next-generation multiple-access (NGMA), simultaneous wireless information and power transfer (SWIPT), and millimeter wave (mmWave). Finally, we outline several research directions for future RIS-aided CF mMIMO systems.
\end{abstract}


\IEEEpeerreviewmaketitle

\section{Introduction}
\subsection{Motivation}
Sixth-generation (6G) networks will be a vital component in all parts of society, industry, and life, given its primary mission to fulfill the communication needs of humans and intelligent machines. Innovative and eye-catching application scenarios will be realized with the advent of 6G networks, including holographic telepresence, e-Health, ubiquitous connectivity in smart environments, massive robotics, three-dimensional massive unmanned mobility, augmented reality, virtual reality, and the Internet of everything \cite{tataria20216g,guo2021enabling}. In the last few decades, several new promising technologies for 6G networks have been introduced including millimeter-wave communications (mWave), massive multiple-input multiple-output (mMIMO), and network densification. They are to provide more effective and efficient wireless communications than ever with an unprecedented increase in data rates, massive connectivity, high-reliability, and low-latency \cite{zhang2020prospective,zheng2023mobile}. Among them, mMIMO is particularly appealing due to its ability to provide excellent quality of service (QoS), such as a 10 times peak rate increase, to a large number of users in the network \cite{larsson2014massive}.

Conventional mMIMO systems consist of a centralized base station (BS) in one cell with a massive array of antennas to serve a smaller number of users simultaneously using the same time-frequency resources \cite{marzetta2016fundamentals}. Furthermore, in various scenarios, the network throughput of mMIMO systems can approach the Shannon capacity via simple linear processing techniques such as maximum ratio (MR) or zero-forcing (ZF) processing \cite{larsson2014massive}. Additionally, as BS antennas are installed in a compact array, mMIMO systems have low-backhaul requirements \cite{bjornson2017massive}. However, traditional cellular networks have a significant problem due to inter-cell interference and path loss, resulting in a deteriorated system performance for users located at the cell edge \cite{ngo2017cell,yang2023bu}.

\begin{figure}[t]
\centering
\includegraphics[scale=0.48]{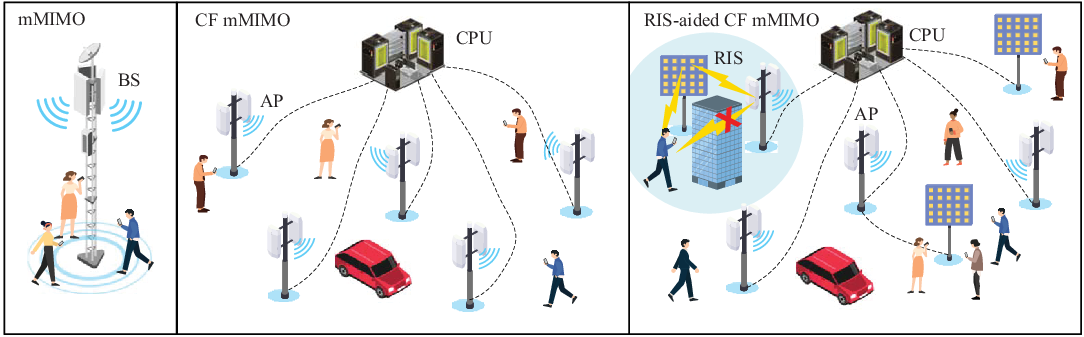}
\caption{Network model evolution. The wireless communication network has evolved from a traditional BS-centric cellular network to a user-centric CF mMIMO network and is further progressing into a more intelligent RIS-aided CF mMIMO network.
\label{Fig1}}
\end{figure}

Different from co-located mMIMO systems, cell-free (CF) mMIMO has been introduced as a promising technology for realizing 6G networks \cite{demir2021foundations}. As depicted in Fig.~\ref{Fig1}, CF mMIMO is a network architecture that comprises numerous access points (APs) that are distributed geographically, connected to a central processing unit (CPU), and collectively serve all user equipments (UEs) via spatial multiplexing on identical time-frequency resources \cite{interdonato2019scalability}. Recently, several important aspects and fundamentals of CF mMIMO have been explored \cite{bjornson2019making,zhang2019cell,elhoushy2021cell,nayebi2017precoding}. The results revealed that the CF mMIMO system can achieve more satisfactory performance compared with the small-cell (SC) system in terms of the cellular edge user throughput \cite{chen2020structured}. Besides, by adopting a distributed architecture and multi-layer signal processing, the system can provide a more uniform QoS, especially for users at the edge of the coverage area \cite{papazafeiropoulos2020performance}. Despite various advantages and potentials of CF mMIMO, ensuring an adequate QoS remains a challenge in harsh propagation environments with insufficient scattering or significant signal attenuation resulting from the existence of substantial obstacles \cite{shi2022spatially}. Meanwhile, as the demand for wireless data throughput continues to soar, meeting increasingly stringent QoS requirements necessitates the implementation of additional advanced technologies.

The reconfigurable intelligent surface (RIS) or intelligent reflecting surface (IRS), is an emerging technology that can alter radio waves at the electromagnetic level without the need for complex digital signal processing and active power amplifiers \cite{wu2019towards,long2021active}. Considering the 3GPP 5G initial access process \cite{3GPP}, RIS can enhance signals and dynamically adjust the wireless propagation environment by connecting to the original communication system. Specifically, RIS can be flexibly deployed to assist users experiencing poor channel conditions in improving communication quality, as it is fabricated using low-power and low-cost material technology \cite{liu2021reconfigurable}. Recognized for their attractive characteristics, RIS is considered to be an effective solution for mitigating various challenges in commercial and civilian applications \cite{tang2020wireless}. However, as a passive device, RIS needs to be integrated with other technologies to play a better role. For example, the authors in \cite{ren2022energy,yang2020performance} proposed an RIS-aided unmanned aerial vehicle (UAV) system to enhance the communication quality and energy efficiency of the system.
Also, the performance analysis and beamforming design of RIS-aided mMIMO systems have also been extensively studied to improve the throughput of existing networks \cite{xu2021sum,tang2020mimo,khaleel2021novel}. The results indicate that the performance of traditional communication systems with poor direct path conditions has a 30\% -40\% improvement with RIS.

Both RIS and CF mMIMO have the ability to achieve more uniform wireless coverage. Integrating RIS into the CF mMIMO system can further achieve 6G ubiquitous high-capacity communication coverage. Therefore, recently, the combination of CF mMIMO and RIS as an evolution of mMIMO system has also attracted widespread attention and research in the industry to achieve the vision of 6G \cite{zhang2021joint,shi2022wireless,zhang2021beyond,le2021energy,al2021ris}. As shown in Fig. 1, the RIS-aided CF mMIMO system consists of a CPU, multiple distributed APs, and distributed RISs. All the APs and RISs are connected to the CPU without cell boundaries to serve all users by coherent transmission and reception over the same time-frequency resources through applying spatial multiplexing techniques \cite{gan2022multiple}. However, the deployment of a large number of APs in the CF mMIMO system requires matching wiring for each AP, which inevitably leads to significant expenses and energy consumption. Also, when there are temporary hotspots that require communication enhancement, deploying additional APs is not practical. RIS, as an easy-to-deploy low-energy device, can solve these challenges of CF mMIMO systems and further improve system performance.
Actually, the RIS-aided CF mMIMO system is not simply a combination of the two technologies, but rather a revolutionary approach that introduces innovations in system architecture, protocols, signal processing, channel models, channel estimation, and other aspects. For example, \cite{yang2022channel,ge2022generalized2} indicate that channel estimation became more difficult and estimation strategies are more diverse.
For the system architecture, \cite{shi2022wireless} points out that the RIS-aided CF mMIMO system is a 3.5-layer architecture, which brings more novel multi-layer signal processing mechanisms. In this way, the system architecture can jointly exploit the advantages of CF mMIMO and RIS to further improve communication performance. For instance, in \cite{zhang2021joint}, a joint AP and RIS precoding framework is proposed and proved to be able to greatly improve user QoS. In addition, previous research has highlighted a major potential of this technology in delivering significant system capacity and achieving superior energy efficiency (EE) while outperforming conventional cellular networks \cite{shi2022wireless,chaaya2023ris}. Along with this, by further incorporating emerging technologies such as next-generation multiple-access (NGMA), simultaneous wireless information and power transfer (SWIPT), millimeter wave (mmWave), terahertz (THz), and unmanned aerial vehicles (UAVs), the RIS-aided CF mMIMO systems can lead to further enhancements in system performance, including achievable data rates, reliability, security, and connection density, thereby paving the way for meeting the target requirements of 6G networks.

\subsection{Comparisons and Key Contributions}
To the best of the authors' knowledge, there are currently no full surveys on the RIS-aided CF mMIMO, but there are separate surveys on CF mMIMO \cite{elhoushy2021cell,ammar2021user} and RIS \cite{liu2021reconfigurable,zhou2023survey,aboagye2022ris}.
More specifically, the authors in \cite{elhoushy2021cell}  provide a comprehensive survey of different aspects of the CF mMIMO system from the general system model, the detailed system operation, the limitations towards a practically implemented system to the potential of integrating the system with emerging techniques/technologies. Then, the authors present several pertinent open problems and outline future directions aimed at realizing the full potential of CF mMIMO systems. Note that the authors believe that the RIS-aided CF mMIMO technology is an important direction in the future. Different from \cite{elhoushy2021cell}, the authors in \cite{ammar2021user} investigate the emerging user-centric CF mMIMO network architecture that sets a foundation for future mobile networks. The key challenges in deploying a user-centric CF mMIMO network and solutions for the main hurdles in CF mMIMO communications are proposed.

On the other hand, the authors in \cite{liu2021reconfigurable} review the physics and communication basic principles of RIS from the perspectives of operating principles, performance evaluation, beamforming design, and resource management. Moreover, the authors focus on investigating the advantages and challenges of machine learning in RIS-enhanced wireless networks. The authors in \cite{zhou2023survey} provide a comprehensive survey on optimization techniques for RIS-aided wireless communications, including model-based, heuristic, and machine learning algorithms. Specifically, different objectives and constraints are summarized and introduced using various optimization algorithms such as alternating optimization and successive convex approximation (SCA). Moreover, the authors present state-of-the-art machine learning algorithms and applications towards RISs, i.e., supervised and unsupervised learning, reinforcement learning, federated learning, graph learning, transfer learning, and hierarchical learning-based approaches~\cite{yang2023detfed,zhang2021optimizing}. The authors in \cite{aboagye2022ris} focus on RIS-assisted visible light communication (VLC) systems. The authors propose a thorough exploration of optical RISs and draw comparisons between optical RISs, radio frequency (RF)-RISs, and optical relays. The critical challenges are highlighted, including the design of RIS element orientations, assignment of RIS elements to access points or users, and positioning of RIS arrays.
Also, there is a short overview paper \cite{shi2022wireless} giving details of RIS-aided CF mMIMO systems from wireless energy transfer (WET) perspective. The authors review the opportunities and challenges of WET in RIS-aided CF mMIMO systems. Specifically, the paradigm of RIS-aided CF mMIMO systems for WET is proposed, including its potential application scenarios, system architecture, hardware design, and operating modes.
However, they mainly focus on a single RIS/CF mMIMO technology and design or provide reviews from a particular perspective. They lack a holistic presentation and comparison of the technical aspects and technical tutorials of RIS-aided CF mMIMO systems.

\begin{figure*}[t]
\centering
\includegraphics[scale=0.55]{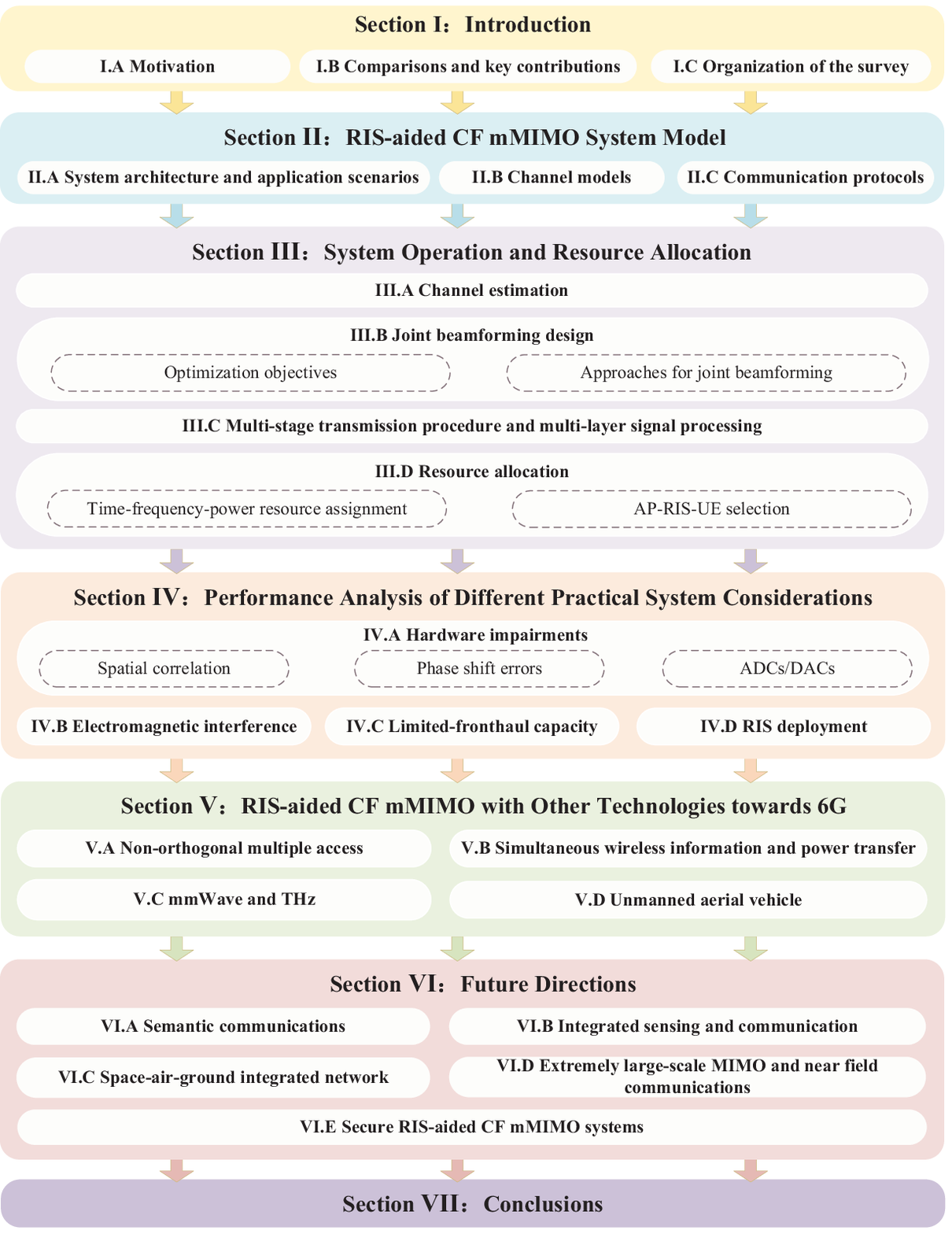}
\caption{The organization structure of the survey.
\label{Fig2}}
\end{figure*}

\begin{table}[ht]
\setlength{\tabcolsep}{2.5pt}
\caption{List of Acronyms \label{Table_1}}
\centering
\begin{tabular}{l l} \hline 
Acronym & Definition \\ \hline
3GPP    & Third generation partnership project   \\
ADC     & Analog-to-digital converter  \\
AP      & Access point   \\
AoA     & Angle-of-arrival   \\
AoD     & Angle-of-departure   \\
BS      & Base station  \\
CB      & Conjugate beamforming \\
CF      & Cell-free  \\
CSI     & Channel state information \\
DAC     & Digital-to-analog converter  \\
DL      & Downlink   \\
FDD     & Frequency-division duplex   \\
HI      & Hardware impairments   \\
HST     & High-speed train \\
IoT     & Internet-of-Thing \\
IRS     & Intelligent reflecting surface \\
ISAC    & Integrated sensing and communication \\
LoS     & Line-of-sight   \\
LS      & Least squares    \\
LSF     & Large-scale fading \\
mMIMO   & Massive multiple-input multiple-output   \\
MMSE    & Minimum mean square error \\
mMTC    & Massive machine type communication \\
mmWave  & Millimeter-wave   \\
MR      & Maximum ratio  \\
NLoS    & Non line-of-sight  \\
NGMA    & Next-generation multiple access \\
NOMA    & Nonorthogonal multiple access \\
OFDM    & Orthogonal frequency-division multiplexing \\
QoS     & Quality of service \\
RIS     & Reconfigurable intelligent surface  \\
RF      & Radio frequency  \\
RL      & Reinforcement learning \\
SAGIN   & Space-air-ground integrated network \\
SWIPT   & \makecell[l]{Simultaneous wireless \\information and power transfer}   \\
SC      & Small cell \\
THz     & Terahertz   \\
TDD     & Time-division duplex \\
UL      & Uplink \\
UE      & User equipment  \\
UAV     & Unmanned aerial vehicle  \\
URLLC   & Ultra-reliable and low-latency communication \\
XL-MIMO & Extremely large-scale MIMO       \\  \hline
\end{tabular}
\end{table}

To this end, we present a comprehensive survey on RIS-aided CF mMIMO systems. More specifically, the goal is to consolidate the state-of-the-art research contributions from the largely fragmented and sparse literature on RIS-aided CF mMIMO systems and highlight future directions. To the best of our knowledge, this survey is the first to comprehensively address the current research status of RIS-aided CF mMIMO across different aspects. The contributions are summarized as follows.
\begin{itemize}
\item We overview RIS-aided CF mMIMO system architecture and major application scenarios in future Internet-of-Things (IoT) networks. We also survey channel models in different propagation environments and provide deep discussions on the main communication protocols.

\item We discuss the system operation and resource allocation, including comparing different channel estimations, and joint beamforming design methods. We also present the unique multi-layer signal processing frameworks of RIS-aided CF mMIMO systems and summarize their advantages and limitations. We highlight and compare the differences between RIS-aided CF mMIMO and the separate, standalone RIS and CF mMIMO. Then, we analyze the importance of resource allocation in RIS-aided CF mMIMO systems and summarize corresponding preliminary research.

\item We investigate the performance of RIS-aided CF mMIMO systems under various practical system considerations, including hardware impairments, electromagnetic interference, limited-fronthaul capacity, and RIS deployment. Notably, we emphasize proposed solutions to address the challenges posed by these practical system characteristics which further facilitates the implementation of RIS-aided CF mMIMO systems.

\item We provide a comprehensive up-to-date review on the combination of RIS-aided CF mMIMO and emerging technologies for 6G networks, including NOMA, SWIPT, mmWave/THz, and UAV. A thorough analysis of the current state-of-the-art and achieved results is provided for each technology. Subsequently, we delve into the limitations of existing works. Finally, we discuss various future directions and identify several open problems that need to be tackled to effectively exploit the potential of RIS-aided CF mMIMO systems.
\end{itemize}

\subsection{Organization of the Survey}
To enhance the coherence of this review, we present Table I which contains the definitions of acronyms used throughout the survey. Subsequently, as illustrated in Fig. 2, the survey is organized as follows. Section \uppercase\expandafter{\romannumeral2} presents the fundamental operating principles of the RIS-aided CF mMIMO system, namely, system architecture and application scenarios, channel model, and communication protocol. Then, Section \uppercase\expandafter{\romannumeral3} focuses on the system operation and resource allocation including channel estimation, joint beamforming, multi-stage transmission procedure and signal processing, and resource allocation. Besides, the system performance analysis of practical system considerations is discussed in Section \uppercase\expandafter{\romannumeral4}. Subsequently, the integration of RIS-aided CF mMIMO with other emerging technologies towards 6G networks is presented in Section \uppercase\expandafter{\romannumeral5}. In Section \uppercase\expandafter{\romannumeral6}, future research directions and several open questions are given. Finally, the conclusions and the key lessons learned in this field are provided in Section \uppercase\expandafter{\romannumeral7}.

\begin{figure*}[t]
\centering
\includegraphics[scale=0.85]{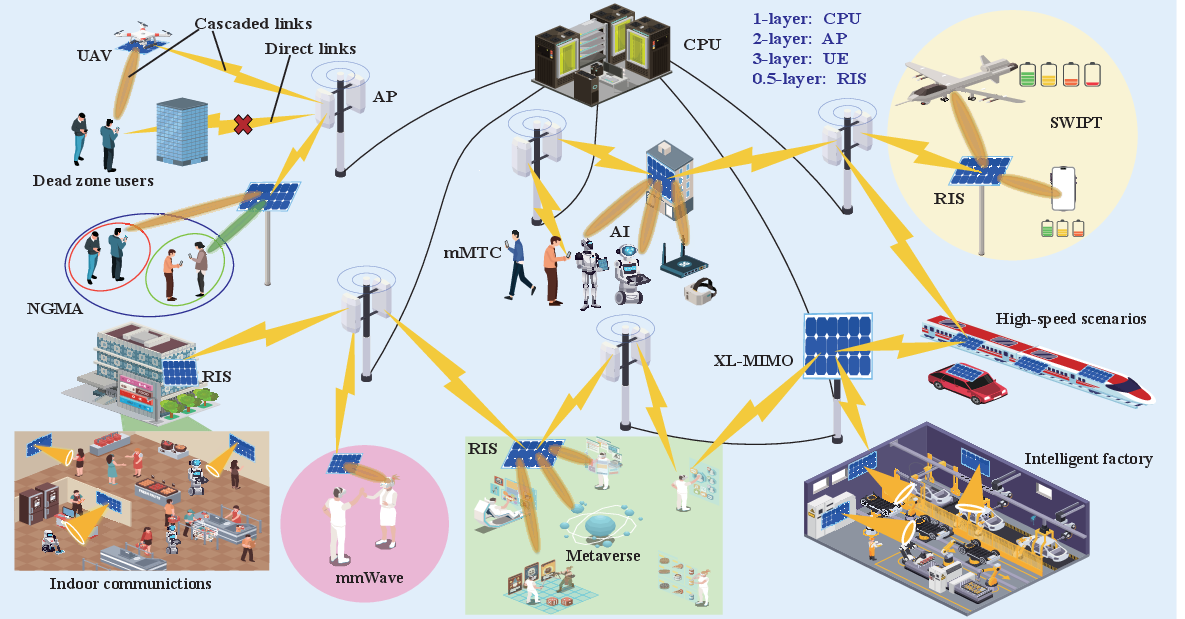}
\caption{Application scenarios of RIS-aided CF mMIMO systems. The scenarios mainly include data demand scenarios such as mMTC, high-mobility, XL-MIMO, mmWave, and Metaverse, as well as energy demand scenarios such as wireless energy transfer in the physical layer.
\label{Fig3}}
\end{figure*}

\section{RIS-aided CF mMIMO System Model}\label{se:model}
In this section, we introduce the system architecture and major application scenarios of RIS-aided CF mMIMO systems. Meanwhile, different channel models among AP, UE, and RIS are compared and analyzed. Moreover, we discuss how the system operates with different communication protocols, such as frequency-division duplex (FDD) and time-division duplex (TDD).

\subsection{System Architecture and Application Scenarios}
As shown in Fig.~\ref{Fig3}, the RIS-aided CF mMIMO system consists of $L$ APs with $M$ antennas, $T$ RISs with $N$ elements, and $K$ UEs with single/multiple antennas that are randomly distributed \cite{chaaya2023ris}. All APs and RISs are connected to the CPU via fronthaul links, which facilitates the exchange of power control coefficients and payload data between the CPU and deployed APs/RISs. Based on that, by utilizing spatial multiplexing, facilitated by a much larger number of APs than UE, the CPU enables all APs and RISs to communicate with all users \cite{elhoushy2021exploiting}. Indeed, RIS-aided CF mMIMO systems are based on CF mMIMO architecture with an ``additional" RIS layer between the UEs and the APs. The CF mMIMO system is structured with three layers consisting of the CPU, APs, and UEs. The RIS-aided CF mMIMO system architecture introduces a 3.5-layer structure by including a cascading link via RISs. The first layer, the second layer, and the third layer are the CPU, the AP, and UE, respectively. In the presence of a direct path, the APs can receive the UE's signal through two uplink paths: the direct link and the aggregated link through RISs. As such, the channels through RIS as an extra 0.5 layer \cite{shi2022wireless}. The RIS layer is referred to as the 0.5 layer because when the direct link is strong, communication can still function properly even without the RIS. In this scenario, the RIS plays a role in auxiliary enhancement. Based on this architecture, the system can ensure stable transmission of information and energy even when the direct path is obstructed.

The future wireless networks are expected to make full use of the low (30 kHz $ \sim $ 300 kHz), medium (300 kHz $ \sim $ 3 MHz), and high ($ \geqslant$ 3 MHz) full-spectrum resources to achieve seamless global coverage such that they can satisfy the stringent demand for establishing unlimited safe and reliable ``human-machine-object" connections anytime and anywhere. Indeed, the success of this desired vision relies on the support of massive access required by the Internet-of-Thing (IoT), requiring higher transmission rates, lower delays, and higher reliability \cite{giordani2020toward,verma2021smart}. Fortunately, RIS-aided CF mMIMO systems can be applied in various application scenarios due to their flexibility and reliability. Fig.~\ref{Fig3} illustrates the applications of the system. First, RIS can be deployed for bypassing the obstacles between APs and UEs, such as adopting a UAV as a carrier to ensure communication connectivity \cite{michailidis2021energy}. In this way, not only can the flexibility of RIS be enhanced, but the use of RF links can be avoided, thereby reducing the energy consumption of UAVs and consequently extending their flight time. Also, RIS as a signal reflection device can support massive machine type communication (mMTC) and intelligent factory via interference mitigation \cite{zhang20196g}. Additionally, RIS-aided CF mMIMO can enhance the electromagnetic signal to serve indoor communications \cite{vasa2022downlink}. On the other hand, RIS-aided CF mMIMO can utilize the distributed architecture in which the RIS compensates for the power loss over long distances with mmWave/THz field and SWIPT networks \cite{ma2023cooperative,yang2021beamforming,khalil2021cure}. Moreover, with the large-scale application of high-speed train (HST) and intercity railways, the demand for high-speed mobile communications is surging. The RIS-aided CF mMIMO architecture can reduce the frequent handover of base stations through distributed architecture and introduce phase compensation to counteract Doppler frequency shift \cite{chaaya2023ris}. Besides, the combination of RIS-aided CF mMIMO and XL-MIMO technology can improve the intelligence of the network and utilize near-field characteristics to improve spatial resolution \cite{xiao2023u,wang2023tutorial}. Moreover, RIS-aided CF mMIMO-enabled ultra-reliable and low-latency communication (URLLC) technology can provide physical layer support for the metaverse \cite{wang2022survey,she2023guest}. To sum up, RIS-aided CF mMIMO systems can provide efficient and reliable physical-layer wireless communication foundational support for various application scenarios.

\subsection{Channel Model}
Different from traditional CF mMIMO channel models which have only one type of links, the RIS-aided CF mMIMO introduces aggregated links through RISs, which are formed by cascading channels between AP-RIS and RIS-UE. As shown in Fig.~\ref{Fig1}, the channels are divided into two types: the direct links from the APs to the UEs and the cascaded links through the RISs. Most of the recent works consider the direct link between UEs and APs as a Rayleigh/Rician fading channel, which assumes non-line-of-sight (NLoS)/line-of-sight (LoS) links between UEs and APs \cite{ngo2017cell,zhang2019cell,bjornson2019making}. Meanwhile, RISs are deployed where there are LoS links between the APs and UEs to enhance communication \cite{wu2019intelligent}. As such, the majority of works utilize the Rician fading to model the channel between AP-RIS and RIS-UE \cite{shi2022uplink,shi2022spatially,nguyen2022spectral}. Besides, some works consider the sparsely-scattered mmWave channel model to describe the channels among APs, UEs, and RISs \cite{ma2023cooperative}. Here, we present a typical case that the UE is equipped with a single antenna, and the channel between AP $l$ and UE $k$ is given by
\begin{align}
{{\mathbf{o}}_{lk}} = {{\mathbf{g}}_{lk}} + \sum\limits_{t = 1}^T {{\mathbf{H}}_{lt}^{\text{H}}{\mathbf{\Phi }_{t}}{{\mathbf{z}}_{tk}}} ,\forall l,k,
\end{align}
where ${{\mathbf{g}}_{lk}} \in {\mathbb{C}^{M \times 1}}$, ${\mathbf{H}}_{lt} \in {\mathbb{C}^{M \times N}}$, and ${{\mathbf{z}}_{tk}} \in {\mathbb{C}^{N \times 1}}$ denote the direct link from AP $l$ to UE $k$, one part of the cascaded channel from AP $l$ to RIS $t$, and another part of the cascading channel from RIS $t$ to UE $k$, respectively. ${{\mathbf{\Phi }}_t} = {\text{diag}}\left( {{e^{j{\varphi _{1t}}}},{e^{j{\varphi _{2t}}}}, \cdots ,{e^{j{\varphi _{Nt}}}}} \right) \in {\mathbb{C}^{N \times N}}$ represents the phase shift matrix of RIS $t$ where ${\varphi _{nt}} \in \left[ { - \pi ,\pi } \right], n \in \left\{ {1, \ldots ,N} \right\},t \in \left\{ {1, \ldots ,T} \right\}$. The direct channel is generally represented as ${{\mathbf{g}}_{lk}} \sim \mathcal{C}\mathcal{N}\left( {{{{\mathbf{\bar g}}}_{lk}},{\beta _{lk}}{{\mathbf{R}}_{lk}}} \right)$ where ${{{{\mathbf{\bar g}}}_{lk}}}$ represents the deterministic LoS component, ${{\mathbf{R}}_{lk}} \in {\mathbb{C}^{M \times M}}$ is the channel correlation matrix of small-scale fading component, and ${{\beta _{lk}}}$ is the large-scale fading (LSF) coefficient which can be obtained from \cite{elhoushy2021cell}.

As for the aggregated channel through RIS, some works consider it as a whole and then model the LSF coefficients ${\beta _{l,{\text{RIS}},k}}$ \cite{tang2022path,tang2020wirelessPath}, while major works consider it as a cascade of two links \cite{van2021reconfigurable,bjornson2020rayleigh}. This is due to the fact that the latter scenario, where there are two cascaded channels, ensures their independence from each other, making it closer to real-world channel conditions. Additionally, during signal processing, it becomes possible to design each of the two-channel segments separately, providing greater flexibility \cite{galappaththige2021performance}. In particular, ${{\mathbf{H}}_{lt}}$ and ${{\mathbf{z}}_{tk}}$ can be denoted as
\begin{align}
{{\mathbf{H}}_{lt}} \sim \mathcal{C}\mathcal{N}\left( {{{{\mathbf{\bar H}}}_{lk}},{{{\mathbf{\tilde R}}}_{lt}}} \right),\,{{\mathbf{z}}_{tk}} \sim \mathcal{C}\mathcal{N}\left( {{{{\mathbf{\bar z}}}_{tk}},{{{\mathbf{\tilde R}}}_{tk}}} \right)
\end{align}
where ${{{{\mathbf{\bar H}}}_{lk}}}$ and ${{{{\mathbf{\bar z}}}_{tk}}}$ denote the LoS components of channel ${{\mathbf{H}}_{lt}}$ and ${{\mathbf{z}}_{tk}}$, respectively. ${{{\mathbf{\tilde R}}}_{lt}} \in {\mathbb{C}^{MN \times MN}}$ and ${{{\mathbf{\tilde R}}}_{tk}} \in {\mathbb{C}^{N \times N}}$ are the spatial covariance matrices which are generated by the joint action of the AP antennas and the RIS elements and will be discussed in Section \ref{se:HI}.

\subsection{Communication Protocol}
The majority of the current works adopt TDD mode as a candidate transmission protocol for RIS-aided CF mMIMO systems, primarily because it is simpler compared to FDD. To the best of our knowledge, there is currently little research on adopting the FDD mode for the systems, while \cite{abdallah2020efficient,kim2018fdd,kim2020downlink} and \cite{guo2021two,zhou2023reconfigurable,chen2021adaptive} have separately studied the application of FDD models in CF and RIS. As illustrated in Fig.~\ref{Fig3}, under the TDD mode, the UL and DL data transmissions take place over the same frequency band, and the frame is divided into three phases in each coherence block: UL training, UL data transmission, and DL data transmission. First, the UEs send $\tau_p$ pilot sequences to APs for channel estimation. Note that, the introduction of RIS brings different operations to the channel estimation stage, such as whether RIS cascaded channels require independent estimation. Also, it is worth noting that during the channel estimation phase, the phase shift of RIS needs to be kept fixed. These are summarized in the subsequent section on channel estimation. Then, utilizing the estimated channels, the vectors required for precoding and detection in the uplink and DL data transmission phases can be computed by APs, respectively. We notice that under the TDD mode, the RIS aggregated channels and the direct channels have reciprocity for uplink and DL communication \cite{galappaththige2021performance}. As such, DL data precoding is performed by each AP using the UL estimate channels and the UEs decode the DL data depending on the channel statistics information that has been obtained by UL training.

By contrast, in FDD systems, the UL and DL data transmissions occur simultaneously, but they are transmitted over two separate frequency bands. As such, the channel coefficients of UL and DL are not reciprocal which requires the APs to acquire the DL estimated channel to perform the DL precoding procedure. Also, as for the RISs, using the same phase shifts will result in beam misalignment, thereby leading to performance degradation \cite{guo2021two,zhou2023reconfigurable}. Indeed, accurate channel estimation can be achieved by adding a DL training phase before DL data transmission. However, this method introduces additional time-frequency resource overhead, which reduces communication rates. Additionally, when there are a number of APs, the exchanged CSI causes a high load to the network \cite{abdallah2020efficient}. To tackle this issue, various CSI acquisition techniques have been proposed by utilizing the angle reciprocity in RIS-aided CF mMIMO systems under the FDD mode \cite{chen2021adaptive,han2022downlink}. However, implementing these methods in RIS-aided CF mMIMO systems requires further investigation.

\begin{figure}[t]
\centering
\includegraphics[scale=0.55]{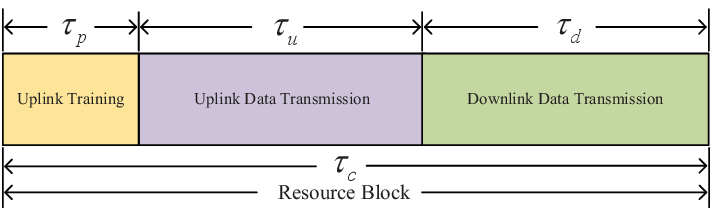}
\caption{Illustration of the considered system frame structure under TDD communication protocol.
\label{Fig4}}
\end{figure}

\section{System Operation and Resource Allocation}\label{se:System Operation and Signal Processing}
In this section, we investigate the system operation of RIS-aided CF mMIMO systems including channel estimation, joint beamforming, and multi-layer signal processing. Besides, we explore the resource allocation approach of the considered system. In particular, regarding the above system operation and resource allocation, we survey the specific issues of RIS-aided CF mMIMO systems and propose several feasible techniques to solve them.

\subsection{Channel Estimation}

\begin{table*}[t!]
  \centering
  \fontsize{9}{12}\selectfont
  \caption{Different Techniques for Channel Estimation in RIS-aided CF mMIMO Systems.}
  \label{System_UPA}
\renewcommand{\arraystretch}{1.5}
\begin{tabular}{!{\vrule width1.2pt}  m{1.8 cm}<{\centering} !{\vrule width1.2pt}  m{2.5 cm}<{\centering} !{\vrule width1.2pt}  m{0.4 cm}<{\centering} !{\vrule width1.2pt}  m{6.5 cm}<{\centering} !{\vrule width1.2pt}  m{5 cm}<{\centering} !{\vrule width1.2pt}  m{1.9 cm}<{\centering}}
   \Xhline{1.2pt}
   \rowcolor{gray!50} \bf Adopted Techniques & \bf Implementation Method & \bf Ref. & \bf Description & \bf Remark \\
   \hline
   \multirow{5}{*}{} & \multirow{3}{*}{}  & \cite{van2021reconfigurable}   &  \makecell[l]{User sends orthogonal pilot signals simultaneously, \\and AP detects the received pilot signals, directly \\estimating the aggregated channel.}  & \vspace{1cm}\multirow{3}{*}{\makecell[l]{The proposed approach does not \\require high training time and pilot \\resources, while being able to capture \\spatial correlation of RIS.}}             \\
   \cline{3-4}
       & \makecell[c]{Aggregated channel \\estimation} & \cite{nguyen2022downlink}   & \makecell[l]{Considering self-interference of RISs, the MMSE \\estimate is adopted to derive the effective channel.}   &              \\
   \cline{3-4}
      \makecell[c]{Pilot-based\\ technique} &  & \cite{vasa2022downlink}     & \makecell[l]{Considering all the users in the same cluster \\transmit the same pilot signal under NOMA \\system, and the LMMSE approach is adopted.}   &             \\
   \cline{2-5}
       & \multirow{3}{*}{\makecell[c]{Separation channel \\estimation}}   & \cite{bashar2020performance}   & \makecell[l]{Divide the training phase into $T\times N+1$ sub-\\phases. First, all RISs are turned off to estimate \\the direct channels, and then switch on each \\element sequentially to estimate the cascading \\channel.}   &  \vspace{-0.5cm}\multirow{3}{*}{\makecell[l]{The proposed scheme requires massive \\pilot sequence resources and training \\time slots, and cannot capture the \\spatial correlation between different \\RIS elements. It is applicable when the \\number of RIS elements is not large. }}  \\
   \cline{3-4}
       &  & \cite{tran2022uplink}   & \makecell[l]{Divide the training phase into $T+1$ sub-phases. \\First, all RISs are turned off and then switched on \\each RIS sequentially to estimate aggregated \\channels.}   &             \\
   \cline{1-5}
       Compressive sensing & Two-timescale channel estimation & \cite{yang2022channel}   & \makecell[l]{A 3D-MMV framework is proposed to jointly \\ estimate cascaded AoDs for all users and utilizes \\tensor contraction to present the 3D-MLAOMP.}   & \makecell[l]{The proposed scheme efficiently reduce \\the NMSE and the performance is \\close to Oracle LS.}            \\
   \cline{1-5}
       \multirow{3}{*}{\makecell[c]{Generalized \\superimposed \\training}} & \multirow{3}{*}{\makecell[c]{Simultaneous \\transmission of \\pilot and data}} & \cite{ge2022generalized1}   & \makecell[l]{The pilots and data symbols are transmitted \\simultaneously in the coherence time. In OFDM \\multi-carrier case, a part of the sub-carriers is \\based on the GST, whereas the other part of \\subcarriers is used for data transmission only.}   & \makecell[l]{The GST scheme can achieve better \\channel estimation performance under \\centralized processing compared with \\local processing.}            \\
   \cline{3-5}
       &  & \cite{ge2022generalized2}   & \makecell[l]{The pilots and data symbols are transmitted \\simultaneously in the coherence time. The channel \\estimation and data detection processes are \\conducted based on the assumption of a certain \\correlation between pilot and data symbols.}   & \makecell[l]{The GST scheme can simultaneously \\reduce NMSE of channel estimation \\compared with the standard ST and \\the regular pilot scheme.}       \\
   \Xhline{1.2pt}

    \end{tabular}
  \vspace{0cm}
\end{table*}

The channel estimation is important for RIS-aided CF mMIMO systems to achieve high spectral and energy efficiencies \cite{yang2022channel,taha2021enabling,chen2023channel,an2022codebook}. The channel estimation directly impacts the calculation of precoding and detection vectors adopted for downlink and uplink transmissions. The problem becomes more complicated with RIS as it is a passive device and cannot independently perform channel estimation itself. If the channel estimation is not accurate enough, it will seriously affect the design of RIS beamforming, leading to deviation in the focusing of the signal beam reflected by RIS. Especially when multiple RISs need to serve multiple APs simultaneously, inaccurate channel estimation can lead to inaccurate beamforming of the APs and phase shift design of the RISs, causing mutual interference signals between multiple RIS and APs to increase, leading to communication performance degradation.

Taking this into account, various techniques are employed in the literature to perform channel estimation in RIS-aided CF mMIMO systems, and the techniques used vary based on the communication protocol adopted, such as TDD or FDD. Currently, most channel estimation studies consider TDD. As such, the pilot-based channel estimation techniques are widely adopted where UEs transmit $\tau_p$-length pilot sequences to the APs as shown in Fig.~\ref{Fig4}. The pilot sequences assigned to UEs can be either orthogonal or non-orthogonal, depending on the number of UEs and the channel coherence time. In general, the UEs can be assigned orthogonal pilot sequences in low mobility scenarios and a small number of UEs. By contrast, in high mobility scenarios with a small $\tau_c$ as mentioned in \cite{zheng2021impact,zhang2021ris}, non-orthogonal pilot sequences are preferred to limit the amount of resources consumed in performing the channel estimation.

Different from traditional CF mMIMO systems, for the aggregated channel introduced by RIS, existing channel estimation studies consider dividing the aggregated channel into two parts, i.e., the direct channel and the cascaded channel, and estimating them separately, namely separation channel estimation \cite{bashar2020performance,tran2022uplink}. In \cite{bashar2020performance,tran2022uplink}, the channel estimation is divided into $1+NT$ sub-phases. In the first phase, all elements of RISs are turned off and each AP estimates the direct channels between itself and all UEs through the pilot transmitted by UEs. In the following $NT$ phases, based on the previously estimated direct channels, the cascaded channels are estimated by sequentially turning on and off each RIS element. Besides, \cite{tran2022uplink} considers estimating the cascaded channel by sequentially turning on and off each RIS. However, the separation channel estimation method requires a large number of pilot sequence resources and training time slots and cannot capture the coupling relationship between different RIS element channels such as the spatial correlation of channels, especially when the number of RIS elements is large. Considering this, researchers have adopted an aggregated channel estimation approach in which the cascaded and direct channels are treated indifferently for channel estimation \cite{van2021reconfigurable,nguyen2022downlink}. For example, in \cite{van2021reconfigurable}, the authors assume that UEs send orthogonal pilot signals simultaneously, and the APs detect the received pilot signals, directly estimating the aggregated channel. The results show that compared to separation channel estimation, this method realizes time savings that are proportional to the number of RIS elements. This suggests that as the number of RIS elements increases, the aggregated channel estimation method becomes more applicable. Also, considering the self-interference of RISs, the MMSE estimate is adopted to derive the effective channel that can reduce the error of channel estimation \cite{nguyen2022downlink}. Besides, in \cite{vasa2022downlink}, the pilot transmission strategy is considered where all UEs in the same cluster transmit the same pilot signal under the NOMA system. Then, the L-MMSE approach is adopted considering the spatial correlation of RIS elements. The aggregated channel estimation approach does not require high training time and pilot resources while capturing spatial correlation of RIS elements at the expense of the computational complexity.

Different from the pilot-based techniques, \cite{yang2022channel} propose two-timescale channel estimation and adopt a compressive sensing technique to improve the accuracy of the estimated channel. In this regard, two distinct characteristics are identified: 1) A common channel between the AP and the RIS for all UEs, and 2) A common channel between the RIS and the UE for all APs. Based on these two characteristics, two-timescale channel estimation issues are investigated which refers to large-scale CSI that undergoes slow transformation and small-scale CSI that undergoes instantaneous changes. Subsequently, two solutions are presented to address these issues: a compressive sensing technique based on a three-dimensional multiple measurement vector (3D-MMV) for cascaded channel estimation and a multi-BS cooperative pilot-reduced methodology for two-timescale channel estimation. The results reveal that the proposed scheme efficiently reduces the NMSE and the performance is close to Oracle LS. Furthermore, the works in \cite{ge2022generalized1,ge2022generalized2} investigate the rate by adopting the generalized superimposed training (GST). In particular, GST involves transmitting pilot and data signals simultaneously within the channel coherence time $\tau_c$, instead of sending them consecutively as shown in Fig.~\ref{Fig4}. The channel estimation and data detection processes are then conducted based on the assumption of a certain correlation between the pilot and data symbols. For instance, \cite{ge2022generalized1} considers the orthogonal frequency-division multiplexing (OFDM) multi-carrier system where a part of the sub-carriers is based on the GST, whereas the other part of subcarriers is used for data transmission only. The results show that applying the GST scheme can reduce the channel estimation normalized means-squared error (NMSE) compared with the standard superimposed training (ST) and the regular pilot scheme.

To the best of our knowledge, compared with the TDD mode, studies on channel estimation of RIS-aided CF mMIMO systems under FDD mode are relatively limited. Fortunately, there are some studies on FDD channel estimation of RIS in other systems \cite{zhou2023reconfigurable,dai2022distributed,ge2023beamforming}. It is worth noting that CSI acquisition and feedback overhead will pose a challenge for channel estimation in FDD-based RIS-aided CF mMIMO systems, as the amount of downlink CSI feedback increases linearly with the number of antennas and APs. To tackle this issue, \cite{dai2022distributed} proposes a path selection-based feedback reduction and partial CSI-based beamforming scheme in RIS-assisted systems. Specifically, the authors assume that the uplink and downlink multi-path components are similar, including the UL angle-of-arrival (AoA) and downlink angle-of-departure (AoD) as well as the LSF coefficients. Then, a dominating path gain information (DPGI) estimation and feedback scheme is proposed, where both the length of downlink pilot signals and the size of the feedback vector are reduced to the number of selected dominant paths. The results show that the SE of the downlink is improved by updating the active and passive beamformers. Besides, in \cite{dai2022distributed}, to enable reliable downlink channel estimation under FDD in the RIS-aided mMIMO system, the authors propose to leverage the distributed machine learning (DML) technique. In particular, the network architecture consists of a downlink channel estimation neural network shared by all users, which can be collaboratively trained by the BS and users using the DML technique. The DML-based hierarchical neural network further improves the accuracy by extracting different channel features. Simulation results indicate that the proposed approach achieves better channel estimation performance and reduces pilot overhead for all users.

\subsection{Joint Beamforming Design}

As described in the previous section, RISs can adjust the phase of incoming electromagnetic waves to effectively improve network performance. However, as a passive device, if proper beamforming design is not carried out, the beam reflected by the RIS will become interference signals in the electromagnetic propagation environment, leading to a decrease in the quality of the expected signal detection and causing degradation in user performance \cite{zhang2021beyond}. In particular, in RIS-aided CF mMIMO systems where multiple APs exist, how to design the phase shift for RISs to provide collaborative services to multiple APs at the same time is an important challenge. Here, we review recent research contributions on the joint beamforming design in RIS-aided CF mMIMO systems. Along the literature review, comparisons are provided among different objectives and methods for facilitating the joint beamforming design, accompanied by their benefits and limitations.

\subsubsection{Optimization Objectives}
In RIS-aided CF mMIMO systems, different application requirements will lead to different optimization objectives, which subsequently involve optimization problems and constraints. In the following, the related research works are reviewed according to the optimization objectives.

\begin{table*}[t!]
  \centering
  \fontsize{9}{12}\selectfont
  \caption{Joint Transmit and RIS Passive Beamforming Design.}
  \label{System_UPA}
\renewcommand{\arraystretch}{1}
\begin{tabular}{!{\vrule width1.2pt}  m{2.5 cm}<{\centering}
!{\vrule width1.2pt}  m{0.8 cm}<{\centering}
!{\vrule width1.2pt}  m{1 cm}<{\centering}
!{\vrule width1.2pt}  m{1.8 cm}<{\centering}
!{\vrule width1.2pt}  m{1.8 cm}<{\centering}
!{\vrule width1.2pt}  m{3.5 cm}<{\centering}
!{\vrule width1.2pt}  m{3.5 cm}<{\centering}
!{\vrule width1.2pt}  m{3 cm}<{\centering}}
   \Xhline{1.2pt}
   \rowcolor{gray!50} \bf Objectives & \bf Ref. & \bf UL/DL & \bf CSI & \bf Phase Shifts &  \bf AP/UE beamforming & \bf RIS beamforming \\ \hline
   \multirow{6}{*}{max WSR} & \cite{zhang2021joint} & DL & perfect & continuous &  PDS & PDS \\
   \cline{2-7} & \cite{xu2023algorithm} & DL & perfect & continuous & DL & DL \\
   \cline{2-7} & \cite{ma2023cooperative} & DL & perfect & continuous & MO & PDS \\
   \cline{2-7} & \cite{huang2020decentralized} & DL & perfect & discrete & ADMM & MM \\
   \cline{2-7} & \cite{gan2022multiple} & DL & two-timescale & continuous & PDS & PDD \\
   \cline{2-7} & \cite{yang2021beamforming} & DL & perfect & continuous & SCA & ADMM \\ \hline
   \multirow{6}{*}{max sum rate} & \cite{xie2020multiple} & DL & imperfect & continuous &  Lagrangian dual sub-gradient & SDR and QCR \\
   \cline{2-7} &\cite{zhang2020reconfigurable} & DL & perfect & discrete & Water-filling & SRO \\
   \cline{2-7} &\cite{huang2021towards} & DL & perfect & continuous & DCP & CGD \\
   \cline{2-7} &\cite{yao2022robust} & DL & imperfect & continuous & SDP & ADMM \\
   \cline{2-7} &\cite{dai2022two} & UL & statistical CSI & continuous & MRC & GA \\
   \cline{2-7} &\cite{cui2022drl} & DL & imperfect & continuous & DRL & DRL \\ \hline
   \multirow{2}{*}{maxmin rate} &\cite{zappone2022achievable} & UL & imperfect & continuous & GP & Alternating maximization \\
   \cline{2-7} &\cite{noh2022cell} & DL & two-timescale & continuous &  ZF & SDR \\ \hline
   \multirow{4}{*}{max EE} &\cite{li2022distributed} & UL & statistical CSI & continuous & DDPG & DDPG \\
   \cline{2-7} &\cite{le2021energy} & DL & perfect & continuous &  IA & IA \\
   \cline{2-7} &\cite{lyu2023energy} & DL & perfect & continuous &  SCA & SCA \\
   \cline{2-7} &\cite{siddiqi2022energy} & DL & perfect & continuous & ZF & SP \\ \hline
   \multirow{2}{*}{max EEF} &\cite{wang2023energy} & UL & perfect & continuous & FP & FP \\
   \cline{2-7} &\cite{liu2021energy} & UL & perfect & discrete & KKT & Lagrangian dual and quadratic transform \\ \hline
   min information leakage &\cite{elhoushy2021exploiting} & DL & statistical CSI & continuous & ZF & SDP \\ \hline
   maxmin SEE &\cite{hao2022robust} & DL & imperfect & continuous & CCCP & SDP \\ \hline
   max WSSR &\cite{hao2022securing} & DL & imperfect & discrete & SDR & SCA \\ \hline
   min network-wide adaption gap &\cite{weinberger2022ris} & DL & perfect & continuous & SCA & SCA \\ \hline

   \Xhline{1.2pt}
    \end{tabular}
  \vspace{0cm}
\end{table*}

\begin{itemize}
\item \textsl{\textbf{Sum rate maximization:}} Focusing on network capacity, we usually consider the problem of maximizing the sum rate. The RIS-aided CF mMIMO system is composed of a large number of distributed independent APs with different positions, making it difficult to obtain real-time and accurate channel information. Therefore, there are more optimization constraints, resulting in greater difficulty and complexity in optimization. For example, in \cite{zhang2021joint}, the authors consider differences in user importance and AP selection and attempt to maximize the weighted sum rate (WSR). By developing an AO framework, they decouple the joint precoding problem and alternately solve the AP beamforming and RIS phase shift subproblems. Specifically, the RIS passive beamforming and AP precoding are designed by the primal-dual subgradient (PDS) approach. The results show that deploying RIS closer to users can achieve better performance than deploying RIS closer to a single AP. Also, by optimizing the phase shift of RIS, when RIS moved from a position 60 meters away from the UEs cluster to a position 10 meters away, the sum rate increased by more than twice. Moreover, the WSR maximization problem is investigated in \cite{xu2023algorithm} which adopts the algorithm unrolling-based distributed optimization. Specifically, they solve this problem by deep distributed alternating direction method of multipliers ($\rm{D}^2$-ADMM) which is a mono-directional information exchange strategy with small signaling overhead. Unlike the AO algorithm in \cite{zhang2021joint}, which focuses on optimizing augmented Lagrangian functions, the ADMM algorithm involves alternating updates of raw variables and Lagrangian multipliers. This makes ADMM perform better on large-scale RIS-aided CF mMIMO systems. Moreover, \cite{yang2021beamforming} considers multiple energy receivers in the system and adopts the SCA algorithm at AP precoding and ADMM at the RIS beamforming to solve the SWIPT problem. The results show that as the number of AP antennas increases, the performance of only optimizing power gradually surpasses that of only optimizing the phase shift of RIS. It reveals that as the number of AP antennas increases, power optimization becomes relatively more important compared to phase shift optimization of RIS. Furthermore, as the AP maximum transmit power increases, the equal power transmission scheme exhibits deteriorated performance owing to the restricted design flexibility in allocating power among the RIS. In \cite{huang2021towards}, the joint beamforming optimization problem is formulated with the objective of maximizing the aggregate throughput while ensuring the outage constraint of the users. A complex gradient descent (CGD)-based algorithm is proposed for the phase shift control, which addresses the unit-modulus constraint and ensures that the aggregate throughput increases monotonically in each iteration. Then, the authors designed a difference of convex programming (DCP)-based algorithm for AP beamforming optimization. The results show that the proposed algorithm achieves an aggregate throughput that is 53.8\% and 25.1\% higher than the cellular MIMO system with zero-forcing beamformer and the RIS-aided CF mMIMO system with random phase shift control.

In contrast to the studies that considered ideal conditions mentioned above, some authors have considered more practical scenarios, such as imperfect CSI or discrete phase shift conditions \cite{huang2020decentralized,zhang2020reconfigurable,yao2022robust,xie2022robust,gan2022multiple}. For instance, \cite{huang2020decentralized} considers the discrete phase shift of RIS, and based on that, jointly designs beamforming for AP and RIS to maximize the weight sum rate. Specifically, the authors adopt the ADMM algorithm and propose that it is not necessary to exchange all CSI among APs in each iteration. The results show that the quantization level of phase shifts is related to the size of RISs and the number of antennas. In \cite{zhang2020reconfigurable}, the authors design the digital beamforming and discrete RIS phase shift by water-filling and iterative algorithm. The results show that when the discrete phase shift is based on 5-bit quantization, the system performance is almost consistent with the continuous phase shift. On the other hand, \cite{yao2022robust} considers the imperfect CSI with an error term and adopts the block coordinate descent (BCD) method to solve the joint beamforming design problem. Indeed, aiming to maximize the worst-case system sum rate, the semi-definite programs (SDPs)-based AP precoding and the ADMM-based RIS beamforming are designed. In \cite{xie2022robust}, the authors consider both channel estimation error and discrete phase shift and adopt the ASO method to solve the RIS beamforming problem. The results show that the RIS can achieve a robust performance against the CSI uncertainty in CF mMIMO systems. By contrast, \cite{gan2022multiple} proposes the practical two-timescale CSI framework, i.e., statistical CSI of RIS beamforming and instantaneous CSI for AP precoding. Then, the authors introduce the penalty dual subgradient (PDS) and penalty dual decomposition (PDD) to solve the joint beamforming design. In the two-timescale framework, the phase shift design of RIS utilizes statistical CSI rather than instantaneous CSI, leading to reduced information exchange frequency, alleviated fronthual link load, and more effective savings on time-frequency resources.

\item \textsl{\textbf{User fairness:}} Ensuring user fairness is an important measure to ensure the reliability of a communication system. It can prevent a large accumulation of resources among high-performance top-tier users, thereby avoiding resource wastage and enhancing overall user satisfaction~\cite{du2023attention}. Generally, user fairness is achieved by maximizing the minimum user rate. Ensuring that each user receives the minimum possible data rate may require introducing complex constraint conditions, which could involve interrelationships among multiple users, thereby increasing the number and complexity of constraints in the problem. Many studies have attempted to address this issue. For example, in \cite{dai2022two}, the authors focus on the Rician channel and propose a design for the RISs passive beamforming using long-time statistical CSI, while adopting the MR combining technique for the APs beamforming based on instantaneous CSI. They then derive closed-form expressions for the uplink achievable rate and optimize the phase shifts of the RISs using a genetic algorithm (GA) to maximize the minimum user rate. The results demonstrate the effectiveness of the two proposed two-timescale schemes and demonstrate that RIS achieves a minimum user rate improvement of 1.5 times in the CF mMIMO system. Moreover, \cite{zappone2022achievable} maximizes the minimum SINR of the RIS-aided CF mMIMO system, where the imperfect CSI and discrete phase shift of RIS are considered. In particular, the receiver filter design is formulated as a generalized eigenvalue problem leading to a closed-form solution, and the RIS phase shifts are designed using an alternating maximization algorithm. It reveals that utilizing statistical CSI has the potential to outperform the scheme based on instantaneous CSI for a moderate to large number of RIS elements, as it reduces the channel estimation overhead. \cite{noh2022cell} proposes a novel two-step algorithm that the long-term passive beamformers using semidefinite relaxation (SDR) at RISs and short-term active zero-forcing (ZF) precoders and long-term power allocation at APs. Notably, the approach can reduce computational complexity and signaling overhead. \cite{jin2022ris} considers single-user and multi-user scenarios with practical discrete phase shifts and aims to maximize the minimum achievable rate. Specifically, the integer linear program (ILP)-based algorithm and ZF-based successive refinement algorithm are adopted for the single-user and multi-user scenarios, respectively. With the proposed algorithm, the minimum achievable rate of the RIS-aided CF mMIMO system is significantly increased, and using 2-bit discrete phase shifts can practically achieve the same performance as continuous phase shifts.

\item \textsl{\textbf{EE and EEF maximization:}} With an increase in the number of antennas, energy consumption has become an important issue that constrains the development of RIS-aided CF mMIMO systems \cite{tataria20216g}. Therefore, many works have tried to find technologies and methods to improve energy efficiency (EE) \cite{li2022distributed,wang2023energy}. \cite{zhang2021beyond} investigates a hybrid beamforming (HBF) scheme consisting of the RIS-based analog beamforming and the digital beamforming at APs to maximize the EE for the downlink. The iterative algorithm is designed to solve this problem and the results show that the considered system has a better EE performance than those of traditional ones including conventional distributed antenna system (DAS). The same problem is further investigated in \cite{le2021energy} by taking into account the limited fronthaul capacity constraints. To solve this problem, the authors introduce the inner approximation (IA) approach. Besides, the authors in \cite{lyu2023energy} explore an RIS-aided CF mMIMO system utilizing hybrid RISs comprising a combination of active and passive elements that can amplify and reflect the incoming signal, respectively. To maximize the EE of the system, a BCD-based algorithm is proposed to decouple variables, and then the SCA method is used to iteratively address the non-convexity of the subproblems. The results show that the proposed hybrid RIS schemes can attain 92\% of the sum rate while achieving 188\% of the energy efficiency of purely active RIS schemes. Different from the above solutions, the authors in \cite{li2022distributed} propose machine learning to solve the problem of maximizing long-term EE. In practice, considering the statistical CSI, the distributed novel hybrid deep deterministic policy gradient (DDPG) framework is adopted to reduce outage probability and enhance robustness. It reveals that the proposed hybrid DDPG-based algorithm has a faster convergence speed and requires less computational resources and communication overhead during the model training compared with the centralized algorithm.

As an extension of maximizing EE, some researchers consider the issue of energy efficiency fairness (EEF) \cite{wang2023energy,liu2021energy}. Specifically, in \cite{liu2021energy}, the authors formulate the precoding design problem to maximize the EE of the worst user in a wide-band RIS-aided CF network. To solve the above problem, an iterative precoding algorithm is proposed to design the subcarrier assignment, power allocation, combining, and precoding by adopting Lagrangian transform and fractional programming (FP). Moreover, the authors in \cite{wang2023energy} investigate the EEF maximization problem with active RISs in the CF mMIMO system and introduce a joint beamforming and resource allocation (JBRA) algorithm. These two studies indicate that both active and passive RISs, with well-designed beamforming, can improve the worst user EE by 13.8\% and 50\%, respectively. However, even with random phase shift, active RIS can still achieve a 40\% improvement in worst user EE.

\item \textsl{\textbf{Information security:}} In \cite{wu2019towards}, the authors highlight that the utilization of RIS, through the beamforming techniques, can effectively safeguard physical layer security by thwarting the unauthorized interception of desired information by potential eavesdroppers. However, in CF mMIMO systems, the presence of multiple APs causes vulnerability for eavesdroppers to intercept information from various sources. Designing RIS to ensure information security poses greater challenges in such scenarios~\cite{zhang2021physical}. To solve this problem, the authors in \cite{elhoushy2021exploiting} investigate the potential of RIS in boosting the secrecy capacities of CF mMIMO systems under spoofing attacks. Indeed, the authors jointly design the RIS phase shifts and AP power coefficients in the downlink to minimize the information leakage to eavesdroppers while maintaining a certain SE for legitimate users. Specifically, they consider that the APs only have the statistical CSI and apply the ZF precoding for data transmission along with SDP for RIS phase shift design. The results reveal that only two RIS panels activated can significantly improve the secrecy capacity and boost the robustness against the higher power of spoofing pilot attacks. In addition, the authors in \cite{hao2022robust} formulate a max-min secure energy efficiency (SEE) problem and then propose the constrained convex-convex procedure (CCCP) and SDP techniques for AP precoding and RIS beamforming. The results show that continuously increasing the number of AP antennas and transmission power is not the optimal choice, and the SEE performance is the best when the transmission power is 15dBm with 3 antennas. In \cite{hao2022securing}, the authors aim to max weighted sum secrecy rate (WSSR) by jointly optimizing the active precoding at the APs and passive beamforming at the RISs with imperfect CSI and discrete phase shift. Especially, due to the CSI of RIS being difficult to obtain, a scheme for optimizing RIS matching UE based on channel state information is proposed, and LCR relaxation constraints are used to transform this problem into an SDP problem for solution. The results indicate that compared to traditional RIS-aided CF mMIMO systems, matching three UEs per RIS can achieve nearly fully connected performance.
\end{itemize}

\begin{table*}[ht!]
  \centering
  \fontsize{9}{12}\selectfont
  \caption{Approaches for Joint Beamforming Design.}
  \label{Approaches}
\renewcommand{\arraystretch}{2}
\begin{tabular}{!{\vrule width1.2pt}  m{1.5 cm}<{\centering}
!{\vrule width1.2pt}  m{2 cm}<{\centering}
!{\vrule width1.2pt}  m{5.9 cm}<{\centering}
!{\vrule width1.2pt}  m{5.7 cm}<{\centering}
!{\vrule width1.2pt}  m{1.4 cm}<{\centering}
!{\vrule width1.2pt}  m{0.8 cm}<{\centering}
}
   \Xhline{1.2pt}
   \rowcolor{gray!50} \bf Techniques & \bf Approaches & \bf Pros & \bf Cons & \bf Ref.  \\ \hline
   \cline{2-5} & SDP & \makecell[l]{$\bullet$ Relax to convex problem \\$\bullet$ Fast convergence\\$\bullet$ Support complex constraints} & \makecell[l]{$\bullet$ Require rank-one solution construction \\$\bullet$ High computational complexity \\$\bullet$ Not suitable for large-scale problems} & \cite{yao2022robust,noh2022cell,hao2022securing,elhoushy2021exploiting}  \\
   \cline{2-5} \multirow{3}{*}{\makecell[c]{Traditional \\optimization}} & Iterative algorithm & \makecell[l]{$\bullet$ Good complexity-performance tradeoff \\$\bullet$ Handle large-scale problems \\ $\bullet$ Support non-convex problems} & \makecell[l]{$\bullet$ May converge to local optimal solution\\$\bullet$ Slow convergence speed\\$\bullet$ Sensitive to initialization} &  \cite{huang2020decentralized,yang2021beamforming,lyu2023energy,zhang2021beyond} \\
   \cline{2-5} & Subgradient & \makecell[l]{$\bullet$ Nonsmooth optimization problems \\$\bullet$ Low computational complexity} & \makecell[l]{$\bullet$ Need to adjust step size \\$\bullet$ Slow convergence speed} & \cite{zhang2021joint,gan2022multiple,ma2023cooperative}  \\
      \cline{2-5} & Manifold optimization & \makecell[l]{$\bullet$ Suitable for high-dimensional, nonlinear, \\\,\,\,\,\,non Euclidean space problems\\$\bullet$ Fast convergence\\ $\bullet$ High accuracy and efficiency} & \makecell[l]{$\bullet$ High mathematical understanding \\\,\,\,\,\,threshold\\$\bullet$ High computational complexity\\$\bullet$ Highly problem-specific} &  \cite{ma2023cooperative,shtaiwi2023sum,wang2021joint} \\ \hline
   Machine learning & \makecell[c]{Deep/\\Reinforcement\\learning} & \makecell[l]{$\bullet$ Can learn from multidimensional data \\$\bullet$ High accuracy and stability \\$\bullet$ Automatically extract features and patterns} & \vspace{-0.5cm}\makecell[l]{\\$\bullet$ Lack of interpretability and transparency \\$\bullet$ Require large amounts of labeled data \\$\bullet$ May not guarantee global optimality} & \cite{li2022distributed,cui2022drl,cui2023digital}  \\ \hline

   \Xhline{1.2pt}
    \end{tabular}
  \vspace{-0.5cm}
\end{table*}

\subsubsection{Approaches for Joint Beamforming}
Based on the above survey, it can be observed that the current approaches for joint beamforming design in RIS-aided CF mMIMO systems mainly fall into two categories: traditional optimization and machine learning.
For traditional optimization, due to the involvement of non-convex optimization problems with multiple variables, the approach often relies on the AO methods. The advantage of this approach is that the active beamforming design becomes a conventional problem once the passive beamforming vector is determined, which has been extensively studied \cite{nayebi2017precoding}. However, designing the passive beamforming under given transmit beamforming vectors remains a challenging problem. As for machine learning, there is relatively limited research on this direction, with only few works. However, considering the complexity of multi-variable optimization, machine learning holds promise as a potential direction for future exploration and discussion \cite{qin2019deep,luong2019applications,zappone2019wireless}.
In the following, we review the approaches employed in current research contributions for joint beamforming design. Table \ref{Approaches} summarizes the characteristics of the approaches.

\begin{figure*}[t]
\centering
\includegraphics[scale=0.68]{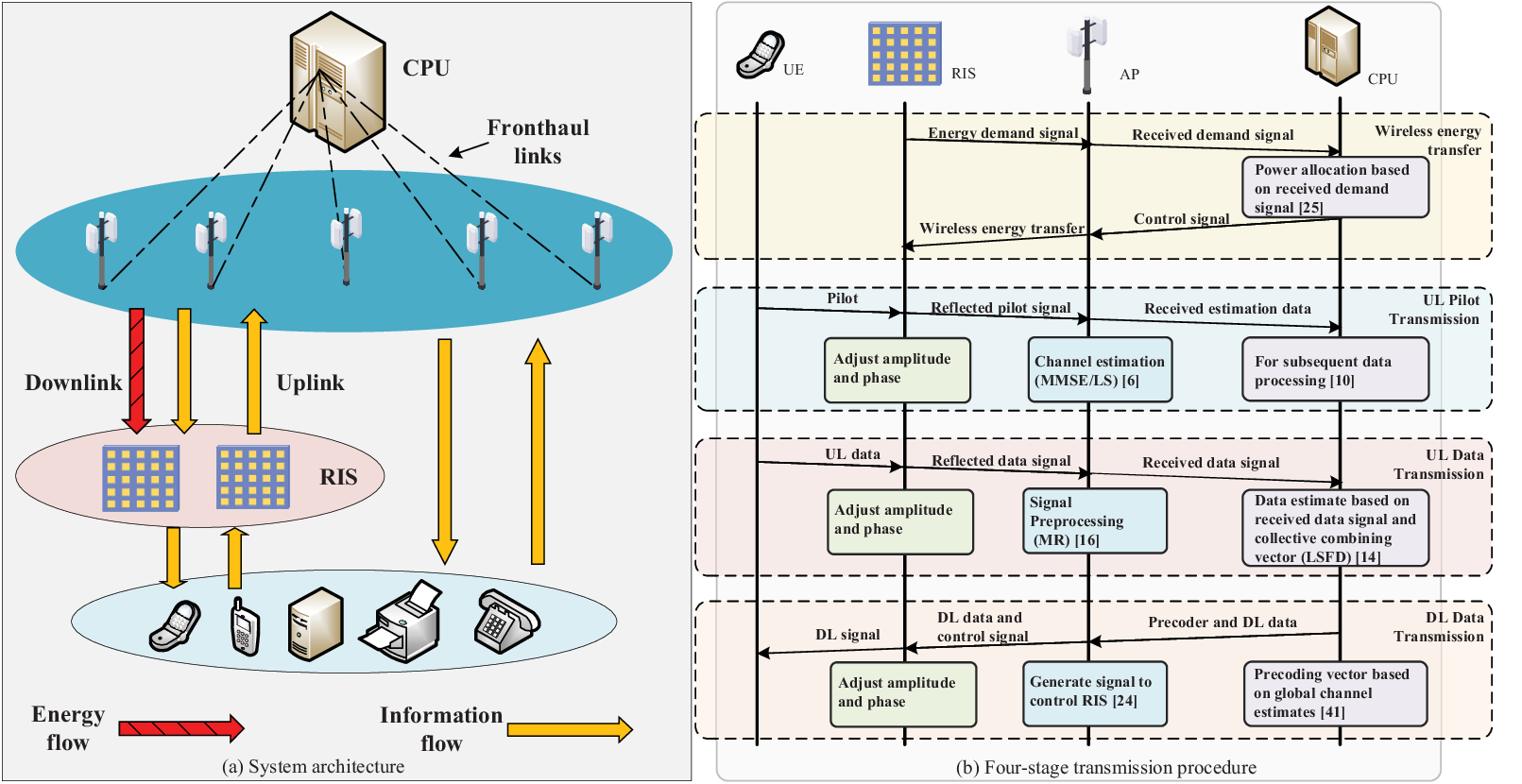}
\caption{Multi-stage transmission procedure and multi-layer signal processing of RIS-aided CF mMIMO systems. (a) Illustration of the system architecture. (b) Illustration of the four-stage transmission procedure, including the wireless energy transfer, UL pilot transmission, UL data transmission, and DL data transmission.
\label{Fig5}}
\end{figure*}

\begin{itemize}
\item \textsl{\textbf{Semi-definite programming (SDP):}} A commonly employed approach for addressing the non-convex unit-modulus constraint involves converting the passive beamforming vector into a rank-one positive semi-definite matrix. Then, the original non-convex problem becomes a convex SDP problem that has fast convergence and supports complex constraints that can be solved by using many efficient convex optimization tools \cite{yao2022robust}. Nevertheless, the rank-one solution constructed in this manner is typically suboptimal and may potentially be infeasible for the original passive beamforming design problem \cite{jin2022ris}. This not only leads to performance degradation but also hinders the convergence of the AO-based iterative algorithm.

\item \textsl{\textbf{Iterative algorithm:}} Iterative algorithm is a commonly used method that strikes a trade-off between complexity and performance, making it suitable for large-scale problems \cite{huang2020decentralized}. Especially, it can also obtain a solution of non-convex optimization problems. However, it may converge to a local optimal solution and have a slow convergence speed \cite{lyu2023energy,zhang2021beyond}. Most importantly, the results are highly sensitive to the initialization.

\item \textsl{\textbf{Sub-gradient:}} The sub-gradient method is a common approach for solving nonsmooth optimization problems. It does not require global derivatives or computations of higher-order derivatives, reducing the computational complexity \cite{zhang2021joint,gan2022multiple,ma2023cooperative}. However, its convergence speed is slow, and the results are highly sensitive to the choice of initialization.

\item \textsl{\textbf{Manifold optimization:}} Manifold optimization (MO) is a common approach for addressing high-dimensional, nonlinear, or problems with specific geometric structures. MO has the characteristics of fast convergence and high accuracy \cite{shtaiwi2023sum}. However, this method has high computational complexity and is only applicable to some specific problems, with low universality.

\item \textsl{\textbf{Machine learning:}} Machine learning can be broadly categorized into several major types, including deep learning, reinforcement learning, and (un)supervised learning. For multi-variable optimization problems, machine learning has distinct advantages \cite{qin2019deep,du2023beyond}. Machine learning can learn features and patterns from complex data, allowing for higher accuracy and stability in the optimization process \cite{li2022distributed}. However, currently, there is a lack of interpretability and transparency in machine learning methods. It often requires a large amount of labeled data to ensure the reliability of results, and cannot guarantee global optimality.
\end{itemize}

\subsection{Multi-stage Transmission Procedure and Multi-layer Signal Processing}
As shown in Fig.~\ref{Fig5} (a), the RIS-aided CF mMIMO system is based on the CF architecture with an additional RIS layer between the APs and the UEs. Different from the existing CF mMIMO system which is a three-tier structure, the RIS-aided CF mMIMO system realizes a 3.5-layer architecture, which adds a cascading link through RIS \cite{shi2022wireless}. Based on the multi-layer nature of the system, performing multi-stage transmission procedures and multi-layer signal processing can reduce inter-user interference and enhance system performance, which is unmatched by traditional cellular networks. In the following, we introduce the multi-layer characteristics of the system in detail, as shown in Fig.~\ref{Fig5} (b).

Stage 1 (Wireless energy transfer): Due to the low power consumption of RIS, some works have proposed wireless energy transfer (WET) technology to further improve the EE of the system \cite{shi2022wireless}. First, the CPU receives information transmitted by the AP via the fronthaul and provides energy control commands to the AP after signal detection. Then, the AP determines whether to transmit wireless energy signals to the RIS for energy harvesting based on the received signals. Specifically, when the energy level stored in the RIS surpasses a predefined threshold, the RIS controller sends feedback signals to the AP to cease the energy transmission.

Stage 2 (Uplink pilot transmission): In the system, each UE is assigned a pilot sequence, which can be allocated randomly or based on a certain metric algorithm \cite{ozdogan2019performance,chen2022improving}. The UE then transmits this pilot signal directly or via the RIS to reach the AP. After receiving the pilot signal, the APs perform local channel estimation, e.g., MMSE and LS channel estimation, to obtain estimated channel information \cite{nguyen2022spectral}. Subsequently, the APs send the required channel estimation data, such as large-scale information, to the CPU for subsequent data transmission processing. Note that during this stage, it is important to keep the phase shift of RISs fixed to ensure the accuracy of channel estimation.

Stage 3 (Uplink data transmission): The UEs send their uplink data to the AP via a direct link as well as the cascaded link through RIS beamforming. Then, the AP executes local signal combining based on the previously estimated channel information. Techniques such as MR/GMR or MMSE combining can be adopted, as proposed in \cite{nguyen2022downlink,GMR}. Subsequently, the local estimates are passed to the CPU for final decoding, where the simple central decoding or LSFD can be employed based on the global channel estimation \cite{bjornson2019making}.

Stage 4 (Downlink data transmission): The downlink data signal is produced by the downlink precoder at the CPU and subsequently transmitted to the AP, allowing it to be conveyed to the UEs through the RIS. In parallel, the AP generates control signals to regulate the phase adaptation of the RIS. It is important to emphasize that in the event that the AP intends to achieve dynamic control over the RIS, adaptations to the frame structure and the inclusion of dedicated control time slots become imperative \cite{chen2022reconfigurable}. Consequently, the deployment of signal processing modules at the RIS may be required to effectively respond to the control signals.

The multi-stage transmission procedure and multi-layer signal processing mechanism are brought about by the inherent distributed structure of the RIS-aided CF mMIMO system. By utilizing the hierarchical signal processing of CPU, AP, and RIS, the performance limits of the system can be further explored while reducing resource costs caused by multi-read information interaction. However, this requires strict protocols and standard guidance to ensure the orderly operation of the system.

\begin{figure*}[t]
\centering
\includegraphics[scale=0.75]{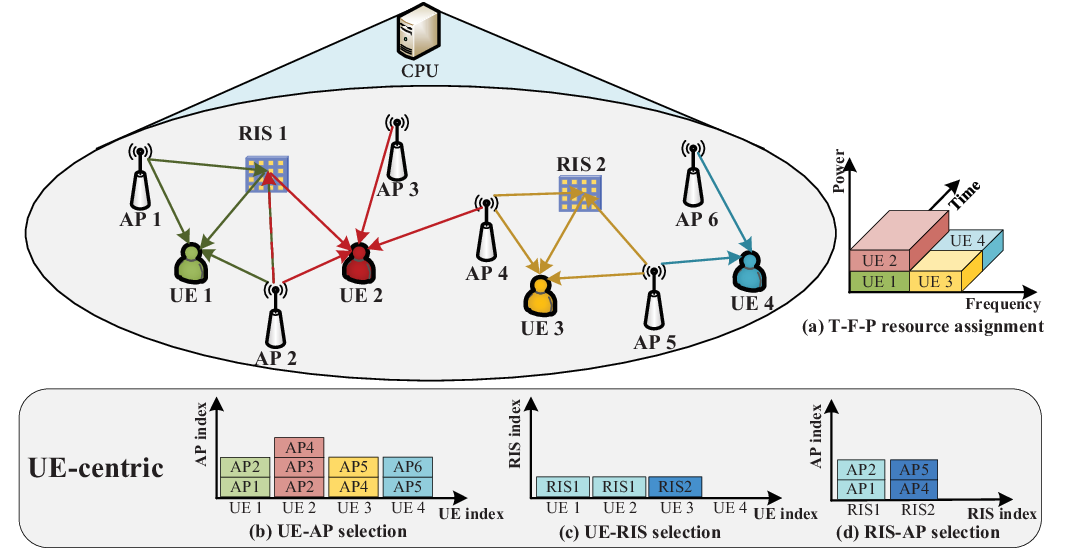}
\caption{Illustration of resource management in UE-centric RIS-aided CF mMIMO systems including the AP-RIS-UE selection and time-frequency-power resource allocation.
\label{Fig6}}
\end{figure*}

\subsection{Resource Allocation}

As shown in Fig.~\ref{Fig6}, in large-scale RIS-aided CF mMIMO networks, proper resource allocation is essential to ensure network stability and efficiency. Specifically, the key difference between RIS-aided CF mMIMO networks and traditional cellular networks is their user-centric nature, where multiple APs and RISs coexist to cater to multiple UEs. In this scenario, addressing time-frequency-power resource assignment and AP-RIS-UE selection is crucial to ensuring their effective coordination and optimizing network performance.

\subsubsection{Time-frequency-power Resource Assignment}
The mainstream research on RIS-aided CF mMIMO systems focuses on system design considering the same time-frequency resources. However, properly assigning users to different sub-channels enhances bandwidth efficiency. In practice, the reflection coefficients of RIS are frequency-dependent. Indeed, when RIS elements lack frequency selectivity, a common RIS reflection matrix must be applied across all sub-channels, leading to challenging optimization problems \cite{cao2021ai,ding2023impact,cao2021reconfigurable}. To solve this difficulty, in \cite{li2022distributed}, the authors propose a DDPG-based algorithm to solve joint beamforming and RIS deployment design for the RIS-aided CF mMIMO NOMA networks to maximize the EE. The introduction of successive interference cancellation (SIC) as an additional optimization constraint in NOMA network optimization has made the optimization problem even more complex. Also, the authors in \cite{vasa2022downlink} derive a closed-form downlink SE expression by considering imperfect CSI and employing imperfect SIC. For clustering, they follow the mechanism where the UEs that have the smallest distance from each other are paired. The results indicate that when AP transmission power is 40 dBm, the incorporation of NOMA results in a 30\% improvement in SE performance compared to OMA. Besides, in \cite{rafieifar2022irs}, the authors design conjugate beamforming for active and passive beamforming at the APs and RISs, respectively. For the RIS assignment, RIS is assigned to the UE with the least distance. Subsequently, they propose a low computational complexity distance-aware UE clustering algorithm where two UEs with the least distance are selected as a pair or cluster. However, this simple grouping approach can also bring many limitations. For example, the spatial distribution of users is often uneven, and users in close proximity may not necessarily have similar channel conditions or requirements. This can lead to a decrease in the performance of some users as they may experience interference or competition. Additionally, the distances and channel conditions between users may change over time with mobility. Therefore, a better approach is to group users based on their channel quality. Grouping users with similar channel quality together can maximize the overall system throughput. These works emphasize different aspects of RIS-aided CF mMIMO systems with NOMA, including joint beamforming and RIS deployment optimization, and UE clustering. However, there are still many research directions that require further investigation, such as dynamic user clustering algorithms, joint time-frequency resource allocation designs, and power allocation problems.

\subsubsection{AP-RIS-UE Selection}
In multi-APs serving UE communication, where there are no longer constraints of cell boundaries, achieving UE-centric communication becomes crucial. To address this, \cite{bjornson2020scalable,chen2020structured} propose a scalable AP-UE selection mechanism to increase the efficiency of CF mMIMO networks. However, an introduction of RIS in CF networks brings forth new challenges in access design, such as UE-RIS and RIS-AP selections. Therefore, the previous AP-UE selection problem has evolved into a joint AP-RIS-UE selection problem. In this context, the optimization problem becomes much more sophisticated.

\begin{itemize}
\item \textsl{UE-AP selection:} Many research studies, such as \cite{chaaya2023ris,xie2022robust,zhou2020achievable}, consider scenarios where all APs serve all UEs. However, this approach significantly increases the computational complexity of the network, especially when a large-scale RIS with a large number of elements is considered. For the UE-AP selection, RIS will change the channel propagation environment and affect user performance, which inevitably leads to differences in the selection results compared to a CF network. Whether the previous UE-centric CF network selection strategies can be extended in the RIS-aided CF mMIMO network is a question to be answered. For example, the authors in \cite{ma2023cooperative} propose a partially-connected CF mMIMO (P-CF-mMIMO) framework to alleviate heavy communication costs. Then, the problem of BS selection is formulated as a binary integer quadratic programming (BIQP) problem, and a relaxed linear approximation algorithm is proposed to address this BIQP problem. The results demonstrate that this optimized UE-AP selection scheme can achieve performance improvement compared with the UE-AP full access scheme while reducing resource consumption. However, there are still many directions that can be explored including how to quantify the impact of RIS during the selection process, designing selection strategies, and optimizing the frame structure.

\item \textsl{UE-RIS selection:} In multi-RIS assisted multi-UE systems, generally the UE-RIS selection schemes determine the overall network performance such that how to associate UEs to different RISs is an important problem. Considering a multi-RIS aided CF mMIMO system, in \cite{bie2023user}, the authors assume that the RISs are deployed near the UE with poor performance and provide one-to-one service for the nearest UE. The results show that when RIS is ${\text{100 m}}$ away from the user, this selection method can achieve a performance improvement of 5 times for the worst performing user compared to without RIS. In addition, the authors in \cite{zhang2021joint} assume that the number of RISs accessed to each UE is limited, and based on this consideration, they formulate the maximum sum-rate optimization problem. For such a zero-one programming problem, the authors employ linear conic relaxation (LCR) to solve it.

\item \textsl{RIS-AP selection:} The energy of the signals reflected by passive RIS is limited and needs to be focused on specific APs in order to achieve significant performance improvements. For simplicity but without loss of generality, the authors in \cite{shi2022spatially} adopt a heuristic distance selection scheme to determine the access of RIS and AP, i.e., each RIS only serves the closest AP. However, getting the distance information, especially for RIS can be challenging in practice. Indeed, in network security scenarios, the selection of RIS and AP is particularly important as it can effectively prevent eavesdroppers \cite{chaaya2023ris,du2022reconfigurable}.
\end{itemize}

\subsubsection{Discussions and Outlook}
In large-scale RIS-aided CF mMIMO networks, resource management is an important direction that has not been fully explored. Based on the above survey, selecting suitable optimization algorithms based on various strategies for AP-RIS-UE selection design is another aspect that needs further research. Possible strategies include channel quality-based grouping strategies \cite{jiang2021grouping}, service quality-based grouping strategies \cite{ma2023cooperative}, and deep learning-based dynamic grouping strategies \cite{gautier2022deep}. Additionally, with the development of computer technology, using machine learning methods to adjust dynamic resource allocation schemes and achieve efficient utilization of time-frequency resources is a promising direction. This will contribute to the orderly and stable operation of the system.

\section{Performance Analysis of Different Practical System Considerations}\label{se:Performance Analysis}
Although RIS-aided CF mMIMO systems have the potential to enhance UEs' performance, several practical limitations may cause severe degradation in system performance. In particular, the integration of RIS and CF will introduce coupling impacts between the inherent practical factors of both, such as spatial correlation and phase shift errors. In this section, we emphasize the impacts of different practical system considerations on system performance and provide insights for the implementation of RIS-aided CF mMIMO systems.

\begin{table*}[ht!]
  \centering
  \fontsize{9}{12}\selectfont
  \caption{RIS-aided CF mMIMO System with Spatial Correlations.}
  \label{Table_HI}
\renewcommand{\arraystretch}{2}
\begin{tabular}{!{\vrule width1.2pt}  m{1.9 cm}<{\centering} !{\vrule width1.2pt}  m{1 cm}<{\centering} !{\vrule width1.2pt}  m{0.5 cm}<{\centering} !{\vrule width1.2pt}  m{2.6 cm}<{\centering} !{\vrule width1.2pt}  m{2.6 cm}<{\centering} !{\vrule width1.2pt}  m{7 cm}<{\centering}|}
   \Xhline{1.2pt}
   \rowcolor{gray!50} \bf \makecell[c]{Hardware\\ impairments}& \bf \makecell[c]{UL/DL}& \bf Ref.& \bf \makecell[c]{RIS element\\ spatial correlation}& \bf \makecell[c]{AP antenna\\ spatial correlation}& \bf Major observation \\
\hline
\multirow{4}*{\bf \makecell[c]{Spatial \\correlation}}&UL/DL&\cite{van2021reconfigurable}&\Checkmark &\XSolidBrush &\multirow{3}*{\makecell[c]{$\bullet$ The spatial correlation of RIS leads to system \\performance degradation.\\ $\bullet$ The system performance degradation caused by \\correlation is minimal at a half wavelength \\spacing of RIS elements.}}\\  
\cline{2-5}  
&UL&\cite{van2021ris}&\Checkmark&\XSolidBrush& \\
\cline{2-5}
&UL&\cite{shi2022spatially}&\Checkmark&\XSolidBrush&\\
\cline{2-6}
&UL&\cite{shi2022uplink}&\Checkmark&\Checkmark&\makecell[c]{$\bullet$ The RIS element correlation in the aggregated \\channel affects the AP antenna correlation, \\resulting in both negative impacts of SE.}\\
\Xhline{1.2pt}
\end{tabular}
\end{table*}

\subsection{Hardware Impairments}\label{se:HI}
RIS is fabricated with low power and low cost which can be flexibly deployed to assist users to improve communication quality. However, this inevitably leads to simple hardware design and limited phase shift accuracy. On the other hand, as the wireless network expands, deploying massive APs will lead to a sharp increase in hardware costs. Therefore, realistic systems tend to deploy low-cost hardware to address cost issues, which can cause degraded system performance. Meanwhile, the hardware impairments of RIS and CF mMIMO systems will couple with each other and cause new impacts. In the following, we enumerate several crucial hardware impairments that significantly affect practical deployments, and subsequently comprehensive investigations to unveil their influence on the system's performance.

\begin{figure}[t]
\centering
\includegraphics[scale=0.5]{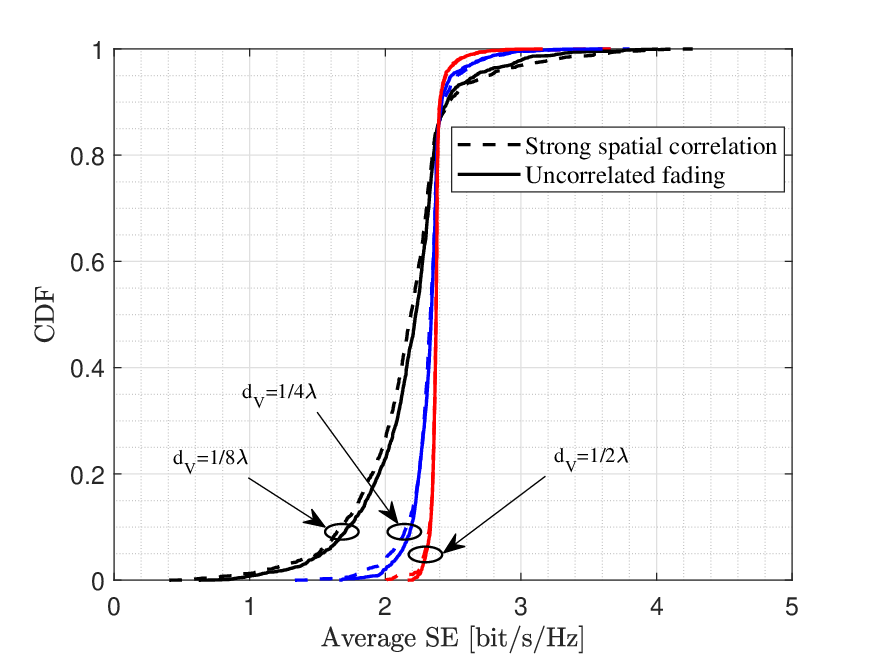}
\caption{CDF of the uplink average SE per UE under different spatial correlations of AP antennas and RIS elements of RIS-aided CF mMIMO systems. The numbers of AP, UE, and RIS element are 40, 10, and 36, respectively. Please refer to \cite{shi2022uplink} for more details.
\label{Fig7}}
\end{figure}

\subsubsection{Spatial Correlation}\label{Spatial correlation}
The presence of spatial correlation in a channel disrupts its inherent characteristics, affecting the accuracy of beamforming and resulting in degraded system performance. In \cite{shi2022spatially}, the authors indicate the existence of two types of spatial correlation in RIS-aided CF mMIMO systems: RIS element spatial correlation and AP antenna spatial correlation. These two types of correlation are mutually coupled, exacerbating the impact on system performance. Table~\ref{Table_HI} summarizes some studies in the literature that analyzed system performance with spatial correlations of AP antennas and RIS elements.

For example, the authors in \cite{van2021reconfigurable,van2021ris,shi2022spatially} consider the impact of RIS element spatial correlation and derive the closed-form expressions of UL and DL SE. In practice, the spatial correlation of RIS elements is modeled by the sinc function introduced by \cite{bjornson2020rayleigh}. The results reveal that the spatial correlation of RIS elements leads to a decrease in system performance gain from the square of the number of RIS elements to a linear increase in the number of RIS elements. Note that at the half-wavelength spacing of RIS elements, the system performance degradation is minimal. Furthermore, the authors in \cite{shi2022uplink} simultaneously consider the spatial correlation of both the RIS elements and AP antennas and derive a closed-form expression for the system average SE utilizing multidimensional matrices. Fig.~\ref{Fig7} illustrates the CDF of average SE per UE under different spatial correlations of AP antennas and RIS elements of RIS-aided CF mMIMO systems. It is clear that the existence of spatial correlations has a negative impact on the system performance, especially the correlation in RIS elements which leads to poor passive beamforming at the RIS.
Note that different from the conclusion that AP antenna spatial correlation is beneficial for CF mMIMO systems in \cite{wang2020uplink}, the RIS elements correlation in the aggregated channel affects the AP antenna's spatial correlation, resulting in both types of correlations having a negative impact on system performance.

\begin{figure}[t]
\centering
\includegraphics[scale=0.5]{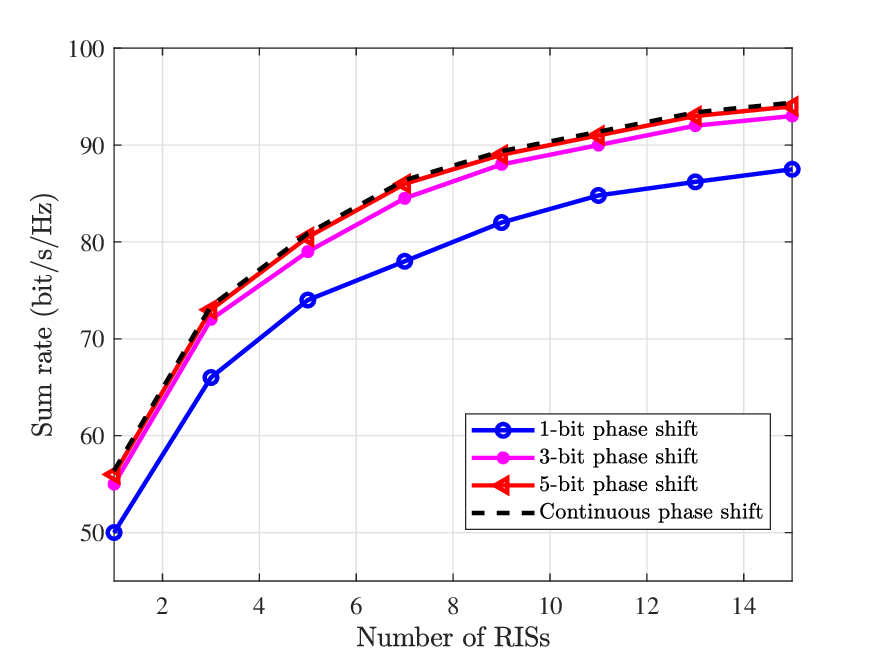}
\caption{Sum rate of the downlink RIS-aided CF mMIMO systems under different phase shift quantization accuracy. The numbers of AP antenna, UE, and RIS element are 8, 8, and 64, respectively. Please refer to \cite{zhang2020reconfigurable} for more details.
\label{Fig8}}
\end{figure}

\subsubsection{Phase Shift Errors}
The primary role of passive RIS in communication systems is to manipulate the phase of the incident electromagnetic waves. Therefore, if the phase of the RIS is inaccurate, it can lead to deviations in beamforming and ultimately affect the signal reception at the AP, thereby impacting the performance of CF mMIMO systems. For this consideration, in \cite{zhang2020reconfigurable}, the authors consider discrete phase shifts of RIS in CF mMIMO systems and quantize the precision using the number of bits. As shown in Fig.~\ref{Fig8}, when using 3-bit quantization, the performance is already close to that of continuous phase shift. Furthermore, the performance obtained using 5-bit quantization is nearly indistinguishable from continuous phase shift. In practice, considering a trade-off between cost and performance, it is common to consider 2-bit and 3-bit quantization precision of RIS elements which has been provided in \cite{zhang2021joint,huang2020decentralized}. It is worth mentioning that the existence of phase shift error introduces additional challenges to the design of joint beamforming optimization algorithms for RIS-aided CF mMIMO systems, especially the discontinuity of RIS phase shift optimization variables. To address this issue, potential solution approaches include approximation algorithms and machine learning techniques \cite{zhakipov2023accurate,sheen2021deep}.

\subsubsection{ADCs/DACs}
\begin{figure}[t]
\centering
\includegraphics[scale=0.5]{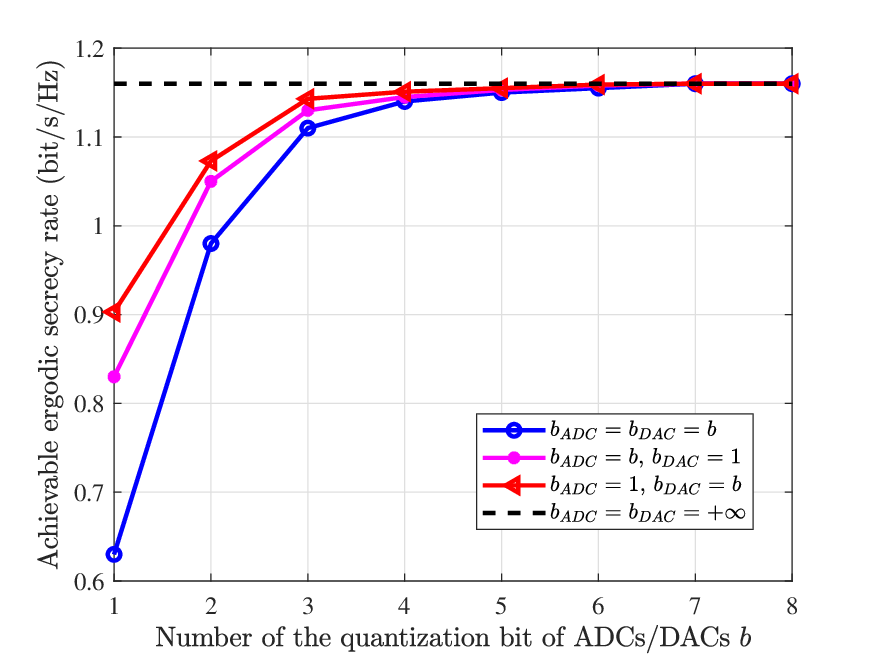}
\caption{The achievable ergodic secrecy rate versus the number of quantization bits of the low resolution ADCs/DACs. The numbers of RIS, RIS element, and AP are 10, 400, and 100, respectively. Please refer to \cite{zhang2022secure} for more details.
\label{Fig9}}
\end{figure}

The presence of numerous distributed APs in networks makes the implementation of ideal high-resolution ADCs/DACs costly and power-intensive \cite{hu2019cell,liu2023cell}. An efficient solution is the utilization of low-resolution ADCs/DACs in cost-effective and energy-efficient CF mMIMO systems. Based on that, in \cite{zhang2022secure}, the authors investigate the impact of low-resolution ADCs/DACs under physical layer security of RIS-aided CF mMIMO systems and derive a closed-form expression of the achievable ergodic secrecy rate. Fig.~\ref{Fig9} illustrates the achievable ergodic secrecy rate versus the number of quantization bits of the low-resolution ADCs/DACs \cite{zhang2022secure}. We can see that 5-bit ADCs/DACs provide sufficiently close results to that of the ideal ADCs/DACs in actual system design.

To sum up, it has been demonstrated that all of the hardware impairments can cause varying degrees of performance degradation in the system. Among them, ADCs/DACs are the most severely damaged as they can simultaneously affect the signals from both the direct link and the RIS cascaded link, resulting in severe distortion of the received signal. However, current research has only focused on the impact of individual factors, lacking a comprehensive consideration of their joint effects. In reality, these hardware impairments coexist and may affect each other. Hence, it is imperative for future research to incorporate a comprehensive consideration of various hardware impairments simultaneously.

\begin{figure}[t]
\centering
\includegraphics[scale=0.5]{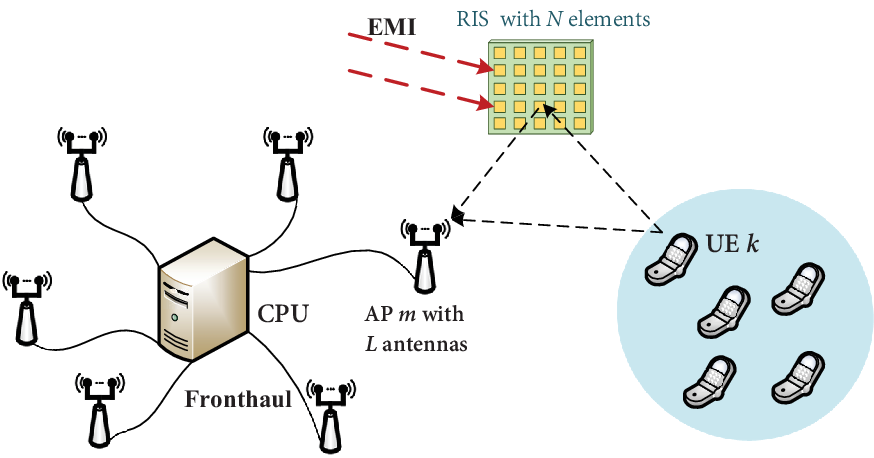}
\caption{An RIS-aided CF mMIMO system with EMI.
\label{Fig10}}
\end{figure}

\renewcommand\arraystretch{1.6}
\begin{table*}[ht]
\caption{RIS-aided CF mMIMO System with Electromagnetic Interference. \label{Table_EMI}}
\centering
\begin{tabular}{!{\vrule width1.2pt}  m{1.5 cm}<{\centering} !{\vrule width1.2pt}  m{3.5 cm}<{\centering} !{\vrule width1.2pt}  m{0.8 cm}<{\centering} !{\vrule width1.2pt}  m{7 cm}<{\centering} !{\vrule width1.2pt}  m{3 cm}<{\centering} |}
   \Xhline{1.2pt}
   \rowcolor{gray!50} \bf \makecell[c]{Target}& \bf \makecell[c]{Scenario}& \bf Ref.& \bf \makecell[c]{Major observation}& \bf \makecell[c]{Future direction}\\
\hline
\multirow{3}*{}&RIS-aided SISO/uplink&\cite{de2021electromagnetic}& \makecell[l]{Due to the presence of EMI, the SNR gain of \\the RIS degrades from $N^2$ to $N$.} & \multirow{3}*{ \makecell[l]{$\bullet$ Joint beamforming \\design}} \\  
\cline{2-4}  
\bf \makecell[c]{EMI}&RIS-aided SISO/downlink&\cite{de2022intelligent}&\makecell[l]{Although RIS aperture can capture much EMI, \\many reflecting elements allow RIS to mitigate\\ EMI effect by means of spatial filtering partially.}& \\
\cline{2-4}
&RIS-aided CF mMIMO/uplink&\cite{shi2023uplink}&\makecell[l]{EMI can degrade system performance, and \\increasing the number of elements can mitigate \\ the impact of EMI.}& \makecell[l]{$\bullet$ Channel estimation}\\
 \Xhline{1.2pt}
\end{tabular}
\end{table*}

\renewcommand\arraystretch{1.5}
\begin{table*}[ht]
\caption{RIS-aided CF mMIMO System with Limited Fronthual Capacity. \label{Table_Fronthual}}
\centering
\begin{tabular}{!{\vrule width1.2pt}  m{1.5 cm}<{\centering} !{\vrule width1.2pt}  m{1 cm}<{\centering} !{\vrule width1.2pt}  m{1.5 cm}<{\centering} !{\vrule width1.2pt}  m{5.5 cm}<{\centering} |}
   \Xhline{1.2pt}
   \rowcolor{gray!50} \bf \makecell[c]{Target}& \bf \makecell[c]{Ref.}& \bf CSI& \bf \makecell[c]{Setting}\\
\hline
\multirow{4}*{\bf \makecell[c]{Limited\\ fronthual}}&\cite{le2021energy}&perfect& \multirow{2}*{\makecell[c]{Fronthual link capacity as a constraint\\ for optimization problems.}} \\  
\cline{2-3}  
&\cite{yao2023robust}&imperfect& \\
\cline{2-4}
&\cite{huang2020decentralized}&perfect&\makecell[c]{Do not have to exchange all CSI\\ among BSs in each iteration.}\\
\cline{2-4}
&\cite{ni2022partially}&perfect&\makecell[c]{Active beamforming vectors are\\ obtained by local APs.}\\
 \Xhline{1.2pt}
\end{tabular}
\end{table*}

\subsection{Electromagnetic Interference}\label{se:EMI}
\begin{figure}[t]
\centering
\includegraphics[scale=0.5]{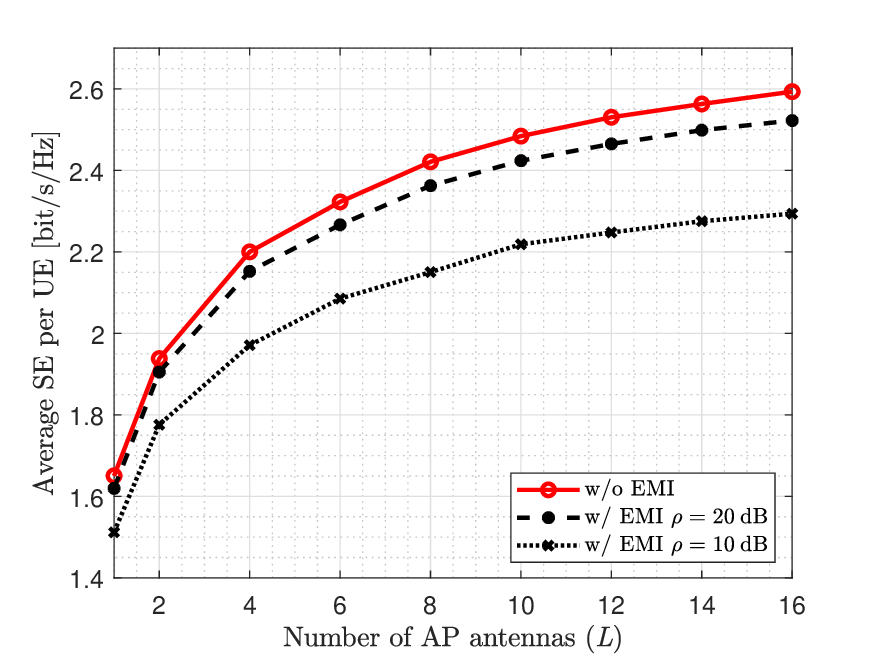}
\caption{Uplink average SE against the number of AP antennas $L$ with different EMI powers with LSFD. The numbers of AP, RIS element, and UE are 10, 16, and 5, respectively. Please refer to \cite{shi2023uplink} for more details.
\label{Fig11}}
\end{figure}

In future wireless propagation environments, spatial electromagnetic interference (EMI) is ubiquitous \cite{de2021electromagnetic}. RISs as passive components, not only reflect useful signals but also reflect EMI signals \cite{de2022intelligent}. This phenomenon introduces inaccuracies in beamforming, subsequently impacting system performance. As shown in Fig.~\ref{Fig10}, in RIS-aided CF mMIMO systems, where the number of APs is substantial, the presence of spatial EMI signals is reflected by RIS. Therefore, when combining RIS with CF mMIMO systems, it is essential to incorporate EMI modeling into the system framework to achieve more accurate performance characterization. Fortunately, in \cite{shi2023uplink}, the authors consider the EMI to further evaluate the RIS-aided CF mMIMO system performance of the actual environment. Then, they derive a closed-form expression for the system SE with the MR combining at the APs and the LSFD at the CPU. Also, the EMI-aware power control methods are proposed to further improve the system performance. As shown in Fig.~\ref{Fig11}, EMI significantly degrades system performance, and the performance gap increases with the increase of $L$ since an increasing number of AP antennas will cause the increased EMI power from RIS. It reveals that when EMI exists, continuously increasing the number of AP antennas in an RIS-aided CF mMIMO system to achieve further system performance improvements is not cost-effective. However, the current research on EMI mainly focuses on performance analysis, and the design of joint beamforming and channel estimation considering EMI scenarios can be a future direction. Table~\ref{Table_EMI} summarizes the related studies and future directions regarding the consideration of EMI in RIS-aided CF mMIMO systems.

\subsection{Limited-fronthaul Capacity}
Unlike centralized architectures, RIS-aided CF mMIMO system performance is greatly limited by the fronthaul link capacity. On one hand, the APs require fronthaul links to connect to the CPU for uplink and downlink data transmission, and the CPU needs to manage power control and beamforming coefficients for different APs \cite{femenias2019cell,masoumi2019performance}. On the other hand, although the RIS is a passive element, it still requires control units for dynamic beamforming design, and obtaining channel information and beamforming instructions from the system requires the fronthaul links \cite{9104346}. As such, in \cite{le2021energy}, the authors focus on maximizing the EE of the RIS-aided CF mMIMO system while considering the constraints of limited fronthaul link capacity. It is assumed that perfect CSI is available for the APs and RISs. Then, the alternating descent-based iterative algorithm is proposed to solve the maximum EE problem. However, obtaining perfect CSI in large-scale networks is challenging. Therefore, in \cite{yao2023robust}, the authors consider robust beamforming design against the impact of imperfect CSI, under the constraint of limited fronthaul capacity. Regarding the optimization subproblem for RIS phase shift, the authors utilize the penalty convex-concave procedure (P-CCP) to attain a stationary solution and establish a robust initialization. As for the AP precoding optimization subproblem, the authors adopt the SCA method, ensuring convergence towards a KKT solution with guaranteed reliability. The results show that when the fronthaul capacity is less than 80 Mbps, the centralized BS demonstrates a superior performance.
Unlike the above works, which consider the fronthaul link capacity as a constraint for global optimization, the works \cite{huang2020decentralized,ni2022partially} take a different approach by considering the exchange of partial channel information to alleviate the capacity requirement of the fronthaul link. For example, the authors in \cite{ni2022partially} consider the active beamforming vectors are obtained by local APs and the passive beamforming vector for RIS is optimized by the CPU so that the fronthaul links do not need to transmit all channel information to the CPU. Table~\ref{Table_Fronthual} summarizes related studies regarding the consideration of limited fronthual capacity in RIS-aided CF mMIMO systems.

However, the above studies assume that the RIS can be controlled instantaneously, without considering the fronthaul link of the RIS. In practice, RIS can be directly connected to the CPU or connected to the local AP for independent phase shift control. For the RIS connected to the CPU, real-time control can be achieved by the CPU conveying processed instructions to the RIS. Conversely, if the RIS is connected to the local AP, the CPU needs to transmit the control signals to the target AP, which decodes the signals and forwards control instructions to the RIS. In this case, the design of the fronthaul link needs to consider an additional fronthaul capacity requirement for the signals generated to the RIS. Currently, the fronthaul link limitation issue of RIS is still open and worth investigating.

\renewcommand\arraystretch{1.5}
\begin{table*}[ht]
\caption{The Deployment of RIS in RIS-aided CF mMIMO Systems.  \label{Table_Deployment}}
\centering
\begin{tabular}{!{\vrule width1.2pt}  m{1.7 cm}<{\centering}
!{\vrule width1.2pt}  m{1 cm}<{\centering}
!{\vrule width1.2pt}  m{1 cm}<{\centering}
!{\vrule width1.2pt}  m{2.5 cm}<{\centering}
!{\vrule width1.2pt}  m{2 cm}<{\centering}
!{\vrule width1.2pt}  m{2.5 cm}<{\centering}
!{\vrule width1.2pt}  m{4 cm}<{\centering} |}
   \Xhline{1.2pt}
   \rowcolor{gray!50} \bf \makecell[c]{Target}& \bf \makecell[c]{Ref.}& \bf \makecell[c]{UL/DL}& \bf \makecell[c]{AP/UE antenna}&\bf \makecell[c]{RIS number}&\bf \makecell[c]{UE distribution}&\bf \makecell[c]{Major observations}\\
\hline
\multirow{4}*{\bf \makecell[c]{RIS\\deployment}}&\cite{zhang2022ris}&DL& \makecell[c]{Single/single}&One&Cluster&\multirow{3}*{}\\  
\cline{2-6}  
&\cite{zhou2020achievable}&DL&\makecell[c]{Single/single}&One&Cluster&\makecell[c]{RISs in close proximity to \\UE clusters leads to \\significant performance.}\\
\cline{2-6}
&\cite{zhang2021joint}&DL&\makecell[c]{Multiple/multiple}&Multiple&Cluster&\\
\cline{2-7}
&\cite{shi2023uplink}&UL&\makecell[c]{Multiple/single}&One&Uniform random distribution&\makecell[c]{RIS in the center of the \\region yields the best \\system performance.}\\
 \Xhline{1.2pt}
\end{tabular}
\end{table*}

\subsection{RIS Deployment}
\begin{figure}[t]
\centering
\includegraphics[scale=0.48]{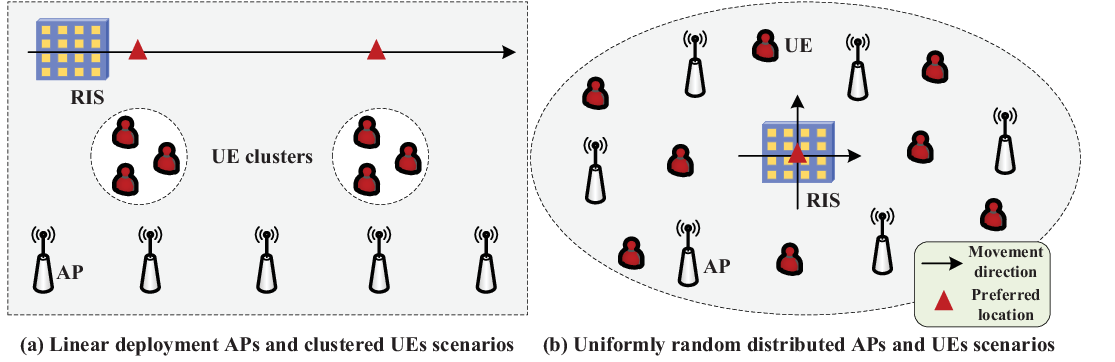}
\caption{RIS deployment for different scenarios of APs and UEs distribution. Two different distribution scenarios are given: (a) Illustration of the preferred location of RIS in linear deployment APs and clustered UEs scenarios. (b) Illustration of the preferred location of RIS in uniformly random distributed APs and UEs scenarios.}\vspace{-0.5cm}
\label{Fig12}
\end{figure}

The path loss of the cascading link via RIS is more severe compared to the direct link. Therefore, the deployment location design of RIS is necessary to achieve desired performance enhancements. Numerous studies have indicated that deploying RIS near the BS or UE can achieve satisfactory performance in BS-centric MIMO systems \cite{liu2020ris,kishk2020exploiting,xu2021reconfigurable,bie2022deployment}. However, in CF mMIMO systems with a large number of distributed APs and RISs, determining the optimal deployment of RISs to achieve the best global performance becomes a challenging problem. The reason is that the optimization problem has to be transformed from a point-to-point scenario to a multi-point-to-multi-point scenario, introducing additional complexities in resource allocation and interference management. We summarize related studies on RIS deployment in RIS-aided CF mMIMO systems in Table~\ref{Table_Deployment}. In \cite{zhang2022ris}, the authors consider maximizing the downlink UE sum rate by alternately optimizing the RIS position and phase shift with perfect CSI. The results show that when UEs are densely distributed, the optimal location for RIS should be closer to the UEs. Furthermore, in \cite{zhou2020achievable}, the joint power allocation, the placement, and the reflection phase shift parameters of the RIS are optimized to maximize the UE achievable rate. However, the above works consider scenarios with only a single RIS and assume that both the APs and UEs are equipped with a single antenna. This simplification significantly reduces the complexity of optimization, but the usefulness is also limited. In practice, RIS-aided CF mMIMO systems involve multiple RISs, UEs, and APs with multiple antennas, which introduces additional challenges in optimizing the deployment of RIS and interference management. Based on that, in \cite{zhang2021joint}, the authors consider the system with multi-RIS and multi-antenna APs/UEs, and then, a joint precoding design scheme at APs and RISs is proposed to maximize the network capacity by adopting a PDS algorithm. The results indicate that deploying RISs in close proximity to UE clusters leads to significant performance improvement. Specifically, in the scenario of two UEs and two APs, RIS improves sum-rate 2.3 times when RIS is 10 meters away from the UE cluster compared to 60 meters away from the UE cluster.
However, in scenarios where APs and UEs are uniformly and randomly distributed throughout the entire system without any UE clustering, the deployment of RIS poses new challenges. Considering this fact, in \cite{shi2023uplink}, the authors consider one RIS serving the CF network where APs and UEs uniformly and randomly distributed within a $1 \,\text{km} \times 1 \,\text{km}$ area. The authors conclude that deploying a single RIS in the center of the region yields the best system performance, achieving an average SE improvement of 7\% compared to deploying RIS at the edge. Based on the above discussion, we illustrate the optimal deployment positions of RISs for different distributions of APs and UEs in Fig.~\ref{Fig12}. To sum up, RIS should be carefully deployed based on the different characteristics of user distribution in RIS-aided CF mMIMO systems.

This section conducts a corresponding survey on the RIS-aided CF mMIMO system from the perspective of practical system considerations and provides some performance analysis results and guiding suggestions. It is worth noting that when the system is launched, it will face many practical problems, such as hardware impairments, limited capacity, etc., and most of these problems will lead to a degradation of system performance. Therefore, we must find the main and secondary factors through reasonable modeling analysis to achieve a tradeoff between performance and cost.

\section{RIS-aided CF mMIMO with Other Enabling Technologies towards 6G}\label{se:Other Technologies towards 6G}
In this section, we provide an overview of the integration of different 6G technologies with RIS-aided CF mMIMO systems. Specifically, we discuss potential benefits of NOMA, SWIPT, UAV, and mmWave technologies with the RIS-aided CF mMIMO systems. Subsequently, we summarize the current research progresses and highlight the existing challenges that require further investigation.

\begin{figure*}[t]
\centering
\includegraphics[scale=0.8]{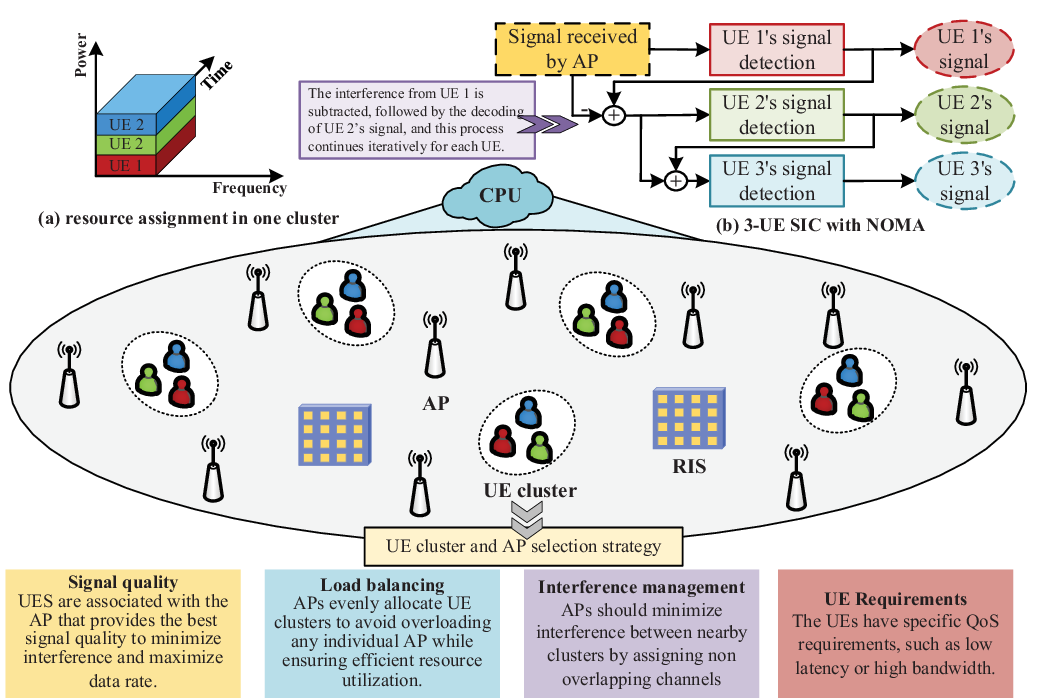}
\caption{An RIS-aided CF mMIMO system with power-domain NOMA where UEs are grouped into clusters. Each cluster includes $l$ UEs, $l = 3$.}\vspace{-4mm}
\label{Fig13}
\end{figure*}

\subsection{NOMA}
The key idea of NOMA is to differentiate signals from different users through different power and coding allocations \cite{saito2013non,zuo2022joint,cao2022massive}. Unlike traditional OMA, NOMA achieves parallel transmission among multiple users by decoding their signals in a non-orthogonal manner. At the receiver, successive interference cancellation (SIC) techniques can be employed to separate and recover the original data signals. As shown in Fig.~\ref{Fig13}, for the RIS-aided CF mMIMO system with power domain NOMA, UEs are grouped into spatial clusters, where the UEs within each cluster share the same time-frequency resources and transmit data symbols with different power levels. Specifically, the AP employs the SIC technique on the received uplink signals that the signals from UE 2 and UE 3 are treated as interference during the decoding process of the signal of UE 1. Then, the interference from UE 1 is subtracted, followed by the decoding of the signal of UE 2, and this process continues iteratively for each UE. Accordingly, NOMA can improve the spectral efficiency, energy efficiency, massive connectivity, and high reliability of RIS-aided CF mMIMO systems \cite{vasa2022downlink,rafieifar2022irs,li2022distributed,zheng2022asynchronous}. In the following, we review the works that considered the UL and DL performance of NOMA-based RIS-aided CF mMIMO systems.

The works in \cite{vasa2022downlink,rafieifar2022irs} examine the DL performance of RIS-aided CF mMIMO systems employing power domain NOMA. In particular, the authors in \cite{vasa2022downlink} investigate the system SE with the imperfect CSI and imperfect SIC at the UEs under spatial correlation RIS. Then, the closed-form DL SE expression is derived by using the linear MMSE channel estimation and statistical CSI. In addition, the authors in \cite{rafieifar2022irs} propose the simple, practical, and independent UE clustering and RIS assignment to UEs algorithms in NOMA. For simplicity but without loss of generality, the conjugate beamforming scheme is adopted for passive and active beamforming at the RISs and APs, respectively. Differently, the authors in \cite{li2022distributed} consider the UL transmission and aim to maximize the EE of the NOMA-based RIS-aided CF mMIMO system. Then, they propose the DDPG-based algorithm to design the joint beamforming for EE enhancement. The key findings in these studies are summarized as follows: 1) The NOMA-based RIS-aided CF mMIMO system operation outperforms the OMA-based RIS-aided CF mMIMO system operation in terms of the number of simultaneously served UEs. 2) Applying NOMA results that the RIS-aided link is more beneficial at lower transmit power regions where the direct link between AP and UE is weak. 3) A small number of RISs increases the ergodic rate. However, a large number of RISs can impact inter and intra-tier interference resulting in a slight decrease in the ergodic rate. 4) Adopting the multi-agent DDPG-based algorithm for the joint beamforming design of RIS and AP can significantly enhance the performance of the considered system.

However, the enhancement of RIS-aided CF mMIMO systems with NOMA relies on optimal clustering of a large number of UEs and making appropriate matching of RIS-assisted UEs and APs. Furthermore, efficient beamforming and power control designs are important to mitigate inter-user interference.

\subsection{SWIPT}
\begin{figure}[t]
\centering
\includegraphics[scale=0.7]{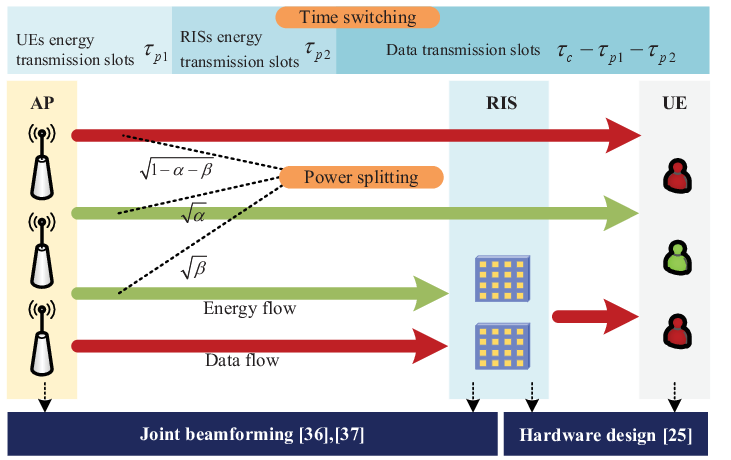}
\caption{SWIPT of RIS-aided CF mMIMO systems includes joint beamforming, hardware design, power splitting technique, and frame structure design.
\label{Fig14}}
\end{figure}

A fast-growing application of IoT resulted in the proliferation of wireless sensors that communicate over the Internet infrastructure~\cite{verma2020network}. Ensuring the long-term operation and self-sustainability of the sensor and IoT devices is a critical consideration in network design. To address this challenge, simultaneous wireless information and power transfer (SWIPT) technology has emerged as a promising solution to extend the lifetime of IoT sensors \cite{huang2018energy,verma2019energy}. CF mMIMO networks have widely adopted SWIPT techniques. Extensive works have demonstrated that the CF network architecture is well-suited for providing energy to UEs that harvest energy, thereby enhancing system-level energy efficiency \cite{demir2020joint,tan2020energy,femenias2021swipt}. Additionally, RIS, as a passive component, exhibits low power consumption. Thus, when wireless energy transfer techniques can be utilized to power the RIS, it can further improve network energy efficiency and eliminate the need for a dedicated power supply for RIS deployment \cite{ren2022energy,yang2021optimal}. Motivated by the aforementioned insights, numerous studies investigated the application of SWIPT techniques in RIS-aided CF mMIMO systems. In the following, we summarize the relevant research findings in this direction as shown in Fig.~\ref{Fig14}.

The work in \cite{yang2021beamforming} investigates the performance of RIS-aided CF mMIMO systems with SWIPT. Specifically, the authors consider that there are multiple information receivers and multiple energy receivers. Based on the assumption that the harvested energy by the UEs should not fall below a threshold, they design a joint beamforming scheme to maximize the achievable sum rate. Then, an SCA-based algorithm is proposed to tackle the energy harvesting constraints, and the ADMM-based algorithm is proposed to iteratively satisfy the norm constraints of RIS reflection elements by applying auxiliary variables and penalty terms. The results show that RIS can effectively enhance the sum rate even without phase shift design. However, increasing the number of APs or transmitting power is not an ideal choice if the APs adopt equal power transmission due to the simultaneous increase in interference signal strength. In \cite{khalil2021cure}, the authors utilize UAVs as mobile APs to provide data communication and energy supply for IoT devices~\cite{verma2017survey}. To achieve the fairness, the max-min rate algorithm is applied to the CPU because the CPU knows the global CSI. The results demonstrate that a uniform deployment of RISs can achieve better performance in terms of energy harvesting. In contrast to the previous works that focus on the joint provision of information and energy services by APs and RISs to UEs, \cite{shi2022wireless} considers the scenario where the APs provide energy to the RISs. As such, the RIS can be deployed without any wired connections. Based on that, the authors propose a four-stage transmission mechanism for SWIPT and introduce three RIS operating modes: centralized RIS, non-cooperative distributed RIS, and cooperative distributed RIS. Also, the authors propose three different RIS hardware design schemes, namely centralized, semi-distributed, and fully distributed, to gradually improve energy reception efficiency.

However, for a large number of APs and RISs, a control protocol must be designed to ensure efficient communication and manage energy interference among devices. For example, as shown in Fig.~\ref{Fig14}, a frame comprises $\tau_p$ number of energy transmission time slots and $\tau_c - \tau_p$ number of data transmission time slots, with $\tau_{p1}$ for UEs energy transmission and $\tau_{p2}$ for RISs energy transmission. Furthermore, in complex networks, factors such as joint beamforming and power splitting are coupled, making the problems non-convex and challenging to solve. Computationally efficient distributed machine learning algorithms serve as attractive solutions, as they have been successfully applied to address large-scale optimization problems \cite{wu2022swipt,zhou2023survey}. Additionally, the hardware design for energy harvesting introduces additional complexity when manufacturing RISs. In fact, designing wireless energy harvesting components at the wavelength level poses significant challenges in controlling cost overhead while ensuring energy harvesting efficiency. Moreover, it is crucial to consider how to avoid coupling interference between power components and information elements at the circuit level.

\subsection{mmWave and THz}
Utilizing mmWave technology in CF mMIMO systems brings significant benefits, including the large bandwidths for high data rates and increased system capacity, as well as the ability to leverage highly directional beamforming and overcome traditional cellular interference, resulting in improved network performance \cite{niu2015survey,du2021millimeter,jin2019channel}. Meanwhile, RIS enables effective beamforming and interference management for mmWave, thereby further enhancing the performance and coverage of mmWave communications \cite{noh2022channel,zarini2023resource,du2022performance}. Therefore, several studies consider the application of mmWave in RIS-aided CF mMIMO systems, summarized as follows.

In \cite{ma2023cooperative}, the authors consider a mmWave communication system with perfect CSI and designed joint beamforming of AP and RIS to maximize the weighted sum rate. Also, the channels including 1 LoS path and 3 NLoS paths are described by the Saleh-Valenzuela model \cite{el2014spatially}. Furthermore, they design an AP-UE partially connected framework to further alleviate the heavy communication overhead in conventional RIS-aided CF mMIMO systems. In addition, the authors in \cite{xu2023algorithm} consider a mmWave channel with perfect CSI, but they employ a distributed cooperation ADMM which adopts a monodirectional information exchange strategy with a small signaling overhead to tackle the joint beamforming problem. The monodirectional information exchange strategy means that all APs have a monodirectional topology, that is, one-way chain propagation of information. The results demonstrate that significant sum-rate improvements can be achieved by employing beamforming designs for mmWave communication in the considered system. However, \cite{ma2023cooperative,xu2023algorithm} all overlook the issue of obtaining the CSI of the mmWave channel. Fortunately, in \cite{lan2023new}, the authors consider deploying an extra AP near the RIS to approximate the mmWave channel from the RIS to the UE and effectively tackle the channel estimation. The results indicate that this approach can achieve system performance close to that with perfect CSI, but deploying a separate AP for RIS channel estimation will introduce additional AP cost overhead.

The aforementioned analysis emphasizes the importance of beamforming in RIS-aided CF mMIMO systems with mmWave. However, in practical implementation, there are several challenges.
First, mmWave channels exhibit highly directional and sparse characteristics, making accurate channel estimation challenging. Proper estimation and tracking techniques are required to effectively utilize passive RISs in mmWave-based RIS-aided CF systems~\cite{zhang2022risp}. Second, deploying RISs in mmWave-based CF mMIMO systems requires careful planning and optimization of the RIS locations. The large number of RIS elements and their precise alignment impose deployment challenges in terms of cost, and practical implementation.

\begin{figure*}[t]
\centering
\includegraphics[scale=0.85]{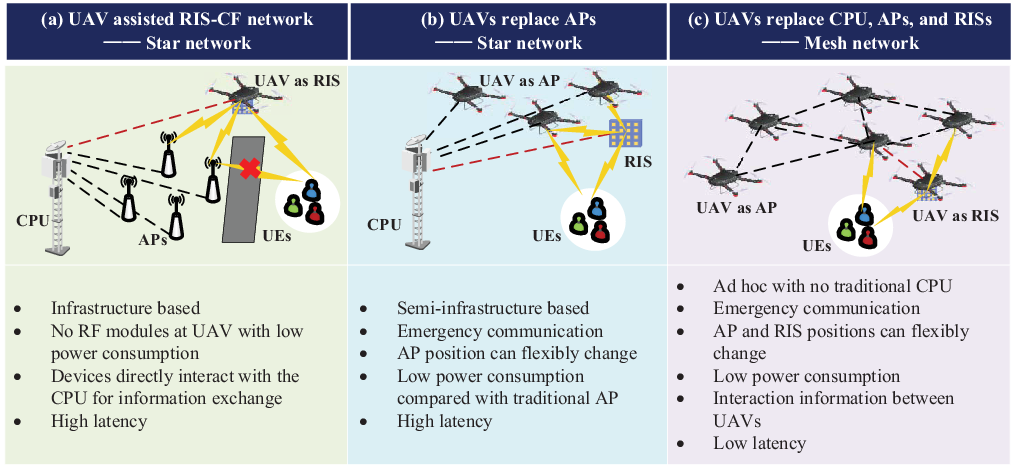}
\caption{UAV with different functions assisted RIS-aided CF mMIMO systems. Two different UAV network modes are provided and their characteristics are analyzed \cite{gupta2015survey}.
\label{Fig15}}
\end{figure*}

\subsection{UAV}
Recently, UAVs have emerged to support diverse applications such as military operations, surveillance, and telecommunications \cite{gupta2015survey,zeng2017energy,du2022performance11}. As illustrated in Fig.~\ref{Fig15}, UAVs function as flying BSs, capable of dynamically adjusting their positions to enhance network coverage and capacity. Additionally, deploying UAVs in CF mMIMO networks benefits users compared to increasing the number of ground-based APs. This advantage arises from UAVs' ability to navigate obstacles effectively, thereby establishing LoS communication links with served users \cite{liu2020cell,zheng2021uav}. In RIS-aided CF mMIMO systems, in addition to serving as mobile APs, UAVs can carry RISs for flexible and mobile deployment of RISs as shown in Fig~\ref{Fig15} (a). This significantly reduces the hardware requirements of UAVs as they no longer need to be equipped with high-energy-consuming RF units, thereby extending their flight endurance \cite{nguyen2022design}. \cite{khalil2023deep} points out that by jointly optimizing the trajectory of UAV and joint beamforming of RIS-aided CF mMIMO systems, the system SE significantly improves. Due to the benefits and potentials, several works have investigated the integration of UAVs with the systems, and our summarized findings are as follows.

To investigate the gain of UAVs in the RIS-aided CF mMIMO system, in \cite{al2022improving}, the authors study the downlink rate based on the assumption that the total power allocated by the CPU to the communication with the UAV and ground UEs is fixed. Specifically, they design CB precoding of RIS and optimize power allocation factors to further improve the downlink rate. The results demonstrate that the combination of UAV and RIS can achieve a 50\% increase in the average system rate, but their effectiveness depends largely on the UAV's height and the number of RIS elements. For example, with 60 elements of RIS, the UAV has a performance improvement of 75\% at a height of 16 meters compared to a height of 100 meters.
The authors in \cite{khalil2021cure} investigate the enhancement of RIS-aided CF mMIMO networks using UAVs and SWIPT technology to improve device energy harvesting efficiency. In particular, they utilize UAVs to replace APs for SWIPT, with each UAV directly controlled by the CPU via fronthaul. The results demonstrate that the deployment of UAV and RIS achieve a 1.5-fold and 1-fold increase in system SE and EE, respectively. Furthermore, compared to centralized and edge RIS deployments, a uniform RIS deployment achieves more desirable performance.
In \cite{khalil2023deep}, an extended study is presented, which introduces a deep learning-based channel estimation framework to eliminate the dependency on traditional closed-form equation-based channel estimation methods, such as LS and MMSE estimation. Additionally, a UAV deployment trajectory planning scheme is proposed to further improve SE. The results indicate that the UAV-empowered RIS-aided CF mMIMO system based on deep learning networks achieves an average SE improvement of 57\% compared to the conventional RIS-aided CF mMIMO system.

In fact, there are different integration schemes for UAVs in RIS-aided CF mMIMO systems, as illustrated in Fig.~\ref{Fig15}. According to the different functionalities of UAVs, the schemes can be classified as UAVs replacing RISs, UAVs replacing APs, and UAVs replacing CPUs. Furthermore, based on different network configurations, they can be further categorized into conventional star networks and mesh networks \cite{cao2021reconfigurableUAV}.
In a star topology, all UAVs are directly connected to the CPU, and all communication between UAVs is routed through the CPU. The star network can achieve global optimal results through centralized optimization and UAV operation at the CPU. However, this configuration can lead to link blockages, increased latency, and a need for high-bandwidth downlinks.
In contrast, mesh networks offer greater flexibility, reliability, and improved performance. In a wireless mesh network, UAV nodes are interconnected and can communicate directly through multiple links. Packets can traverse intermediate nodes, finding their way from any source to any destination through multiple hops. Fully connected wireless networks provide advantages in terms of security and reliability. A promising technique for mesh network configurations is Multi-agent deep reinforcement learning (MADRL) which has been widely applied to multi-node problems and has shown excellent performance \cite{chen2022resource,du2023generative}.
As shown in Fig.~\ref{Fig15} (a), with traditional infrastructure of CF mMIMO systems, the deployment of UAVs equipped with RIS can effectively address coverage gaps when users are located in blind spots. For example, the authors in \cite{yu2022fair} propose a max-min throughput problem in RIS-UAV-enabled mobile vehicle communication. The original problem is broken down into three subproblems: optimizing joint passive beamforming and mobile vehicle scheduling, power allocation, and trajectory. These subproblems are effectively solved using the SCA method and the results show that after adding trajectory optimization, the worst-case user throughput increases by 21\%. This strategy works with the existing network and offers the advantage of RIS mobility. Furthermore, UAVs equipped with RIS have lower power consumption and longer endurance compared to active relay UAVs, as they do not require RF modules.
As illustrated in Fig.~\ref{Fig15} (b), in situations where deploying a wired network is impractical, e.g., for emergency communication, the UAVs can act as alternatives to APs for delivering communication services. For example, in \cite{samir2020leveraging}, the authors utilize the UAVs to provide coverage to vehicles entering a highway that is not covered by other infrastructure. Then, the authors frame the decision-making process for trajectory planning as a Markov decision process (MDP). Subsequently, deep reinforcement learning (DRL) is employed to introduce a method for learning the optimal trajectories of the deployed UAVs. The results show that deploying two UAVs can achieve a 25\% increase in coverage compared to deploying one UAV. However, UAVs necessitate periodic energy recharging to sustain their operation. To tackle this issue, several studies have proposed leveraging wireless energy transfer technology to provide power supply for UAVs \cite{shi2022wireless}.
While the star network relies on a ground-based CPU, as shown in Fig. 15 (c), the mesh network can have UAVs perform the roles of CPU, APs, and RISs. The mesh network enables a decentralized architecture, facilitating seamless communication among UAVs without the need for centralized CPU involvement. By establishing a self-organizing network, the RIS-aided CF mMIMO systems can benefit from UAV-to-UAV information exchange. Moreover, the UAV closest to the RIS can provide control of the RIS phase shift. This novel UAV mesh network enables full-fledged mobility with the utmost flexibility in positioning while maintaining low power consumption and latency. However, due to the lack of a fixed central processing unit, the interaction of global information and the orderly control of UAVs have become the main challenges. Specifically, considering the joint optimization of UVA trajectory, RIS and UAV beamforming, and power control in mobile scenarios is the main technical challenge.

To sum up, there are still many challenges for the UAV-enhanced RIS-aided CF mMIMO systems, such as optimizing UAV trajectory and communication interaction based on partial information. Furthermore, joint beamforming and collaboration between multiple UAVs and multiple RISs are necessary for network intelligence and orderliness.

\section{Future Directions}\label{se:Future Directions}
In this section, we discuss several future research directions for the RIS-aided CF mMIMO system, including semantic communications, integrated sensing and communications (ISAC), space-air-ground integrated network (SAGIN), extremely large-scale MIMO (XL-MIMO) and near field communications, and secure RIS-aided CF mMIMO systems.

\subsection{Semantic Communications}
Embracing the era of 6G wireless communications, it is imperative to explore more advanced technologies that can enhance the intelligence and effectiveness of communication networks. Semantic communications, as an emerging paradigm, is such a technology that considers the semantic information of the transmitted data and optimizes the communication process accordingly~\cite{shi2021semantic}.

RIS-aided CF mMIMO systems can benefit from integrating semantic communications by exploiting semantic information for effective resource allocation~\cite{yang2022semantic}. More specifically, for the RIS-aided CF mMIMO systems, semantic communications can be implemented in the following ways:
\begin{itemize}
    \item In channel estimation and signal processing, a synergistic approach that contemplates semantic information, such as content type and priority, can be effectively employed. Through RIS phase shift optimization and beamforming, this semantic information can be integrated, facilitating communication more tailored to the unique requirements of the data.
    \item In the transmission phase, semantic information about the context, e.g., the location, movement patterns, and data usage behavior of the UEs, can be leveraged for dynamic RIS configuration, enabling an agile response to changing network conditions and user requirements \cite{du2023semantic,zhang2020performance}.
    \item In the resource allocation and networking phase, semantic information transmission scheduling can be implemented. For instance, in a scenario where edge users transmit high-priority or delay-sensitive data, the CF mMIMO network, powered by semantic information, could dynamically allocate more resources to these users~\cite{du2023attention}, ensuring their QoS.
\end{itemize}
The integration of semantic communications with RIS-aided CF mMIMO systems is an area of ongoing exploration. This integration could resolve critical issues and provide symbiotic benefits. For instance, inter-user interference, a notable problem in RIS-aided CF mMIMO systems, could be alleviated by leveraging semantic information and its priority to allocate network resources in an intelligent and dynamic way~\cite{wang2023semantic}.

\subsection{Integrated Sensing and Communication}

ISAC is a novel paradigm aimed at unifying the capabilities of sensing and communication into a single network infrastructure with the same time/frequency resources. The concept of ISAC revolves around the idea of using wireless signals for both data transmission and environment sensing \cite{liu2022integrated}. ISAC plays a crucial role in various application scenarios, such as autonomous driving, smart cities, and IoT \cite{cui2021integrating}. However, the joint design and optimization of sensing and communication functions in a unified system pose significant challenges, primarily due to their different requirements for radio resources~\cite{du2023semantic}.

The integration of ISAC into RIS-aided CF mMIMO systems opens up opportunities for enhancing the performance and functionality of these networks \cite{li2022distributed,cui2023digital}
\begin{itemize}
    \item The distributed nature of CF mMIMO systems makes them an ideal candidate for implementing ISAC. The spatially distributed APs can collaboratively perform sensing tasks and communication services simultaneously, leading to more accurate and extensive environmental perception \cite{demirhan2023cell}.
    \item RIS, with its ability to control the radio propagation environment, can provide hardware-level support for implementing ISAC~\cite{du2023semantic,tran2022uplink}. By intelligently adjusting the phase shifts of the RIS elements, the system can control the direction and power of the reflected signals, thereby enhancing the sensing capabilities of the network \cite{xu2023algorithm,cui2022drl}. Moreover, the RIS can help in mitigating the potential interference between sensing and communication signals, thereby improving the overall network performance \cite{luo2022joint}.
\end{itemize}
However, an integration of ISAC and RIS-aided CF mMIMO systems also introduces new challenges, such as the joint optimization of sensing and CF-aided communication functions and the design of RIS phase shifts for enhancing the sensing capability. Thus, this integration necessitates further research and investigation.

\subsection{Space-Air-Ground Integrated Network}
SAGIN is a promising architecture designed to provide seamless coverage, high data rates, and reliable communication services by integrating space, aerial, and terrestrial networks \cite{liu2018space}. It aims to overcome terrestrial networks' coverage limitations and satellite networks' communication latency issues by leveraging their complementary characteristics. This integrated architecture brings significant benefits, such as expanded coverage area, enhanced network robustness, and improved quality of service \cite{liu2018space}.

Incorporating the RIS-aided CF mMIMO framework into SAGIN introduces new dimensions for improving network performance and system functionality:
\begin{itemize}
    \item CF mMIMO can extend the terrestrial coverage by spatially distributing multiple APs over a large area, thus overcoming the limitations of traditional cellular networks \cite{liu2020cell}. This spatial distribution guarantees consistent service quality throughout the coverage zone and introduces technical challenges and opportunities. Specifically, it necessitates solutions for joint trajectory planning, effective mobility management, and adaptive beamforming.
    \item RIS can enhance both terrestrial and non-terrestrial networks by intelligently controlling the propagation of radio waves, thereby improving the coverage and reducing the energy consumption \cite{bariah2022ris, lin2022refracting}. Moreover, it can aid in the seamless integration of heterogeneous networks by improving inter-network communication and mitigating inter-network interference.
\end{itemize}
However, the efficient and seamless integration of heterogeneous networks poses significant challenges, including network synchronization, efficient resource allocation, and seamless handover management~\cite{sheng2021space}.

\subsection{Extremely Large-scale MIMO and Near Field Communications}

XL-MIMO technology is another remarkable area where RIS-aided CF MIMO systems can bring significant enhancements. XL-MIMO systems, characterized by the deployment of a massive number of antennas, have the potential to boost spectral efficiency, significantly increase data rates, and improve the energy efficiency of wireless communications \cite{wang2022extremely,marinello2020antenna}.
Integrating XL-MIMO into RIS-aided CF mMIMO systems to replace existing APs presents exciting opportunities for enhancing network performance and functionality \cite{lu2021near}
\begin{itemize}
    \item {\textit{Dimensionality and Near-Field Channel Characteristics:}} The increase of antenna numbers in XL-MIMO augments the system's dimensionality, intensifying the inherent complexities of near-field channel characteristics \cite{cui2022near}. These complexities are further exacerbated by the addition of intelligent reflecting surfaces in the RIS-aided CF mMIMO system, which, while enhancing signal coverage, also necessitates an advanced understanding and management of the emerging channel characteristics~\cite{zheng2023flexible}.
    \item {\textit{Signal Focusing and Multipath Fading:}} The substantial antenna array in XL-MIMO provides robust signal-focusing capabilities~\cite{du2020sum}. Yet, the amalgamation of RIS-aided CF mMIMO introduces challenges related to interference management. The utilization of intelligent reflecting surfaces for beamforming certainly aids in reducing interference, but it also amplifies the concerns associated with multipath fading~\cite{demirhan2022enabling,du2019distribution}.
    \item {\textit{Network Adaptability and Configuration:}} Although the distributed architecture of CF facilitates a seemingly fluid integration of XL-MIMO, adapting to communication demands across different scenarios with the ever-evolving RIS technology adds layers of complexity in terms of network configuration, deployment, and real-time adaptability~\cite{kim2020downlink}.
\end{itemize}
Although the combination of XL-MIMO and RIS-aided CF mMIMO technologies holds tremendous potential, it also brings forth various challenges, such as the unique characteristics of near-field channels arising from the substantial increase in XL-MIMO antenna quantity and high signal processing complexities introduced by the ultra-large-dimensional channel matrix between RIS and XL-MIMO \cite{cui2022channel}. Therefore, RIS-aided CF XL-MIMO systems require further research and exploration.

\subsection{Secure RIS-aided CF mMIMO Systems}
In the rapidly evolving of wireless communication, ensuring security against sophisticated eavesdropping techniques is paramount \cite{du2023rethinking,xu2023reconfiguring,du2023spear}. The incorporation of RIS in CF mMIMO systems has emerged as a potential solution to address these challenges~\cite{egashira2022secrecy}:
\begin{itemize}
    \item {\textit{Countering Active Eavesdropping with RIS:}} Active eavesdroppers, often referred to as Eves, present unique challenges in CF mMIMO systems. Specifically, they can exploit pilot contamination attacks, causing increased rates of information leakage. One of the effective ways to mitigate this risk is through the joint optimization of downlink power coefficients at APs and the RIS's phase shifts. This strategy was shown to significantly minimize the risk of information leakage to active Eves, harnessing the unique capabilities of RISs to boost security in CF mMIMO systems \cite{elhoushy2021exploiting}.
    \item {\textit{Balancing Energy Efficiency and Security in RIS-aided CF Networks:}} With an increasing number of legitimate users and eavesdroppers in CF networks, achieving energy efficiency and security becomes a complex optimization problem. Addressing this, researchers have explored the joint design of distributed active beamforming, artificial noise at base stations, and passive beamforming at RIS. Leveraging techniques like fractional programming, this iterative approach has shown promising results in enhancing energy efficiency while ensuring robust security against potential eavesdroppers \cite{hao2022robust}.
    \item {\textit{Harnessing Artificial Noise Against Passive Eavesdropping:}} The distributed nature of CF mMIMO offers spatial diversity benefits, not just to legitimate users but, unfortunately, also to passive eavesdroppers. This dual-edged sword necessitates innovative countermeasures. One such solution is the deployment of an artificial noise (AN)-aided secure power control scheme. This technique exploits spatial diversity for legitimate users and strategically deploys AN to neutralize the gains made by passive eavesdroppers. The outcome is an improved secrecy performance, vital in ensuring secure communications in CF mMIMO networks \cite{park2022secure}.
\end{itemize}
In essence, integrating RIS in CF mMIMO systems promises enhanced communication capabilities and a robust shield against eavesdropping. As these technologies converge, there is a compelling case for further research to design more secure and efficient wireless communication paradigms.

\section{Conclusions}\label{se:conclusion}
6G wireless systems are expected to surpass existing limits and realize a vision of ubiquitous connectivity, ultra-large capacity, ultra-low latency, enhanced coverage, and green communications, thereby enabling the Internet of Everything. Integrating user-centric CF mMIMO and RIS has brought new vitality and potential to support the ambitious goals of 6G wireless networks. In this context, we comprehensively review RIS-aided CF mMIMO systems.
First, we presented the distinctive features of the RIS-aided CF mMIMO system model, particularly highlighting the difference in introducing RIS on CF mMIMO systems. Subsequently, various approaches for channel estimation and joint beamforming design were proposed, and a comprehensive investigation on resource allocation was conducted, providing essential fundamentals and insights for the research on RIS-aided CF mMIMO systems. Also, the unique multi-layer transmission procedure and signal processing mechanisms were emphasized. Then, we underscored the novel practical challenges brought by the integration of RIS in CF mMIMO systems. Insightful simulation results were provided to guide system deployment and practical implementation. Moreover, we summarized the integration of other key technologies with the RIS-aided CF mMIMO system and offered valuable insights into their feasibility. Last, we presented many RIS-aided CF mMIMO-empowered application scenarios and potential future directions.
Our survey serves as a guideline for primary RIS-aided CF mMIMO research works in future 6G communications from the perspective of system modeling, resource allocation and operation, practical performance analysis, the integration of other key technologies, RIS-aided CF mMIMO-empowered application scenarios, and promising future directions.

\bibliographystyle{IEEEtran}
\bibliography{IEEEabrv,Ref}

\end{document}